\documentclass[useAMS,usenatbib]{mn2e}
\usepackage{epsfig}
\usepackage{amsmath}
\title{Long-term Nonlinear Behaviour of the Magnetorotational Instability in a Localised Model of an Accretion Disc}
\author[L. J. Silvers ]{L. J. Silvers\thanks{E-mail:
lara.silvers@lra.ens.fr (LJS)} \\Laboratoire de Radioastronomie,
Departement de Physique Ecole Normale Superieure, 24 Rue Lhmond,
75231, Paris, 05, France.}
\begin{document}

\date{}

\pagerange{\pageref{firstpage}--\pageref{lastpage}} \pubyear{2002}

\maketitle

\label{firstpage}

\begin{abstract}

For more than a decade, the so-called shearing box model has been used
to study the fundamental local dynamics of accretion discs.  This
approach has proved to be very useful because it allows high
resolution and long term studies to be carried out, studies
that would not be possible for a global disc.

Localised disc studies have largely focused on examining the rate of
enhanced transport of angular momentum, essentially a sum of the
Reynolds and Maxwell stresses.  The dominant radial-azimuthal
component of this stress tensor is, in the classic Shakura-Sunayaev
model, expressed as a constant $\alpha$ times the pressure. Previous
studies have estimated $\alpha$ based on a modest number of orbital
times.  Here we use much longer baselines, and perform a cumulative
average for $\alpha$.   Great care must be exercised when trying to extract
numerical $\alpha$ values from simulations: dissipation scales,
computational box aspect ratio, and even numerical algorithms all
affect the result. This study suggests that estimating $\alpha$
becomes more, not less, difficult as computational power increases.

\end{abstract}

\begin{keywords}
accretion discs; magnetic fields; instabilities
\end{keywords}

\section{Introduction}

Linear stability analysis has shown that a differentially rotating
disc, with angular velocity decreasing outwards, is unstable in the presence of a weak magnetic field (see, for
example, \cite{BH91}; \cite{BH92}). This magnetorotational
instability (MRI) gives rise to turbulence within the disc and is a
plausible way that efficient outward transport of momentum in a disc
can be realised (\cite{BH98}).  However, while this is an appealing concept, it is
crucially important, due to the non-linear form of the governing
equations, to investigate the non-linear evolution and saturation of
this instability to determine if this is indeed a route by with
angular momentum transport can be achieved on the required
time-scales.

The equations that govern the evolution within a disc are
computationally demanding to evolve. Therefore, in order to study
the evolution and saturation of key quantities associated with this
instability (e.g.\ stresses, energies etc), it is useful to employ a
localised, `shearing-box' model (see, for example, \cite{B95};
\cite{HGB}). This approach has proved useful as it has enabled
higher local resolutions to be achieved for the same computational
cost as that for a full disc simulation. However, while a localised
approach does facilitate longer runs and higher resolutions, at the
point of the early shearing box calculations the resources were such
that only modest run times (usually less than 30 orbits) were
possible (see, for example \cite{HGB}).

When discussing the MRI, the principal quantity of interest is
$\alpha$, which is related to the rate of angular momentum transport
in a disc. This quantity is defined as (\cite{SS73}; \cite{HGB}):
\begin{equation}
\alpha=w_{xy}/P
\end{equation}
In this definition, P is the gas pressure and $w_{xy}$ is a
component of the $x-y$ stress tensor i.e.\
\begin{equation}
w_{xy}=\rho v_x v_y-\frac{B_x B_y}{4\pi},
\end{equation}
where $\rho$ denotes density; $v_x$ is the $x$ component of the
velocity field; $v_y$ is the perturbed component of the velocity
field in the $y$ direction; $B_x$ is the $x$ component of the
magnetic field and $B_y$ is the $y$ component of the magnetic field.
\footnote{Note here that (x,y) refers to a local set of coordinates
  in (r, $\phi$) and not ($-\phi$,r) as sometimes utilised in certain
  hydrodynamical models.}

In turbulent flows transport quantities such as $\alpha$ are highly
fluctuating. The resource constraints at the time of the earliest
calculations forced $\alpha$ values to be quotes from by a simple
average over a modest number of orbits, which is useful to give an
idea of the order of magnitude of this quantity (see, for example,
\cite{HGB1}). However, unless an extremely large number of
orbits is considered, the saturation level will vary as you increase the number of
orbits over which the average is evaluated. This can, make it difficult to see
clearly the effect if varying each of the parameters (e.g.\ dissipation
scale, domain size etc) associated with the problem.
It is better, if the resources are available, to calculate a cumulative
average for this fluctuating quantity. This does require vast
numbers of computer hours, which is only recently possible. As such,
the principal aim of this paper is to calculate long-time cumulative
averages of $\alpha$.

In order to keep the computational costs acceptable, we choose to
make several simplifying assumptions in this paper. First, we choose to initially
work in a cubic domain, which is only relaxed later in the paper. This
enabled us to explore the effect of decreasing the dissipation scale
on the cumulative average $\alpha$ and to show that small scales are
playing a role in determining the saturation level of this quantity.
The second simplification is that we chose to solve the ideal MHD
equations. This reduces the cost of the calculation by a significant
fraction though it does mean that all calculations are at an
effective Prandtl number of order unity. This is acceptable as in
this particular investigation we were not trying to examine the
effect of varying the Prandtl number. Indeed, this has recently been
considered (\cite{LL2007}; \cite{FP2}). The work in this paper is aimed,
somewhat, to complement these recent works even though here
we do not explicitly include dissipation coefficients.  We will show
that, as in the \cite{LL2007} paper the effect of increasing
resolution for a fixed box size is to increase $\alpha$. Further, we
show that the effect of increasing the box length is to decrease the
saturation level of $\alpha$ as the parasitic instability is allowed
to take hold.

The paper will proceed as follows: In section 2 we will fully detail
our model and the numerical method used in the solution of the
governing equations. We will discuss the results in section 3. This
section is divided into two subsections. In section 3.1 we begin by
showing the problems that are encountered in standard shearing box
calculations that makes long-term cumulative averages difficult to
calculate. We show that this problem is a result of compounded errors
that come from the shearing boundary conditions. We then present a
correction to the numerical algorithm, which permits us to calculate
long-term cumulative averages and show how $\alpha$ varies with
resolution (or dissipation length scale). In section 3.2 we bring the
results section to a close by showing that the effect of increasing
the length of the box in the azimuthal direction is to decrease
$\alpha$. In Section 5 we summarise the finding as well as
discuss the astrophysical consequences.

\section[]{The Model}

In this paper we restrict consideration to a local patch of a
Keplerian disc and to use the shearing box approximation. Within
this paper we choose to follow the strategy of many of the earlier
works by which we mean that we will consider an ideal MHD problem.
We shall assume for simplicity in this paper that the $z$-component
of gravity is suitably small and so may be neglected (assumed to be
within a pressure scale height of the midplane of the disc). Further
we assume that the gas is isothermal. As such, equation set may be
written as:
\begin{equation}
\frac{\partial \rho}{\partial t}+ \nabla.{\rho \mathbf{v}}=0
\label{eqnstart}
\end{equation}
\begin{eqnarray}
\frac{\partial \mathbf{v}}{\partial t}+\mathbf{v}.\nabla \mathbf{v}
=   &- &\frac{1}{\rho}\nabla\left(P+ \frac{B^2}{8\pi}\right) +
\frac{\mathbf{B}.\nabla \mathbf{B}}{4\pi\rho}   - 2 \mathbf{\Omega}
\times \mathbf{v} \nonumber \\ &+ &2q\mathbf{\Omega}^2x \mathbf{\hat{x}}
\end{eqnarray}
\begin{equation}
\frac{\partial \mathbf{B}}{\partial t}=\nabla \times (\textbf{v}
\times \mathbf{B})
\end{equation}
\begin{equation}
\nabla. \mathbf{B}=0
\end{equation}
\begin{equation}
P=c_s^2 \rho \label{eqnend}
\end{equation}
where we obey the standard notational convection and thus, $\rho$
denotes density, $P$ denotes the gas pressure, $v$ and $B$ are the
velocity and magnetic fields, respectively, $\Omega$ is the angular
velocity, $q=-d \ln \Omega/d \ln R$.

Our work will differ from earlier works in the following ways. First
we choose to start from a perturbation that has a specific analytic
form that is not a random perturbation at each of the grid points.
This initial perturbation in such simulations are immaterial so long
as they are small in amplitude. However, in order to be completely
clear on our starting condition, we choose to start from a
perturbation that can be expressed in a simple analytic manner so
that we have a consistent start condition as we vary the resolution.
We choose to perturb each of the velocity components in the
following manner:

Defining
\begin{equation}
f_x=\sum_{n=1}^5  \sin[2 \pi n {x-(1.2-0.2n)}],
\end{equation}
\begin{equation}
f_y=\sum_{n=1}^5 \sin[2 \pi n {y-(1-0.2n)}],
\end{equation}
\begin{equation}
f_z=\sum_{n=1}^5 \sin[2 \pi n {z-(0.8-0.2n)}]
\end{equation}
the initial velocity components are then expressible as:
\begin{equation}
v_x=0.01\Omega f_xf_yf_z
\end{equation}
\begin{equation}
v_y=-q\Omega x + 0.01\Omega f_xf_yf_z
\end{equation}
\begin{equation}
v_z=0.01\Omega f_xf_yf_z
\end{equation}
We have chosen this particular form to give the perturbation some
complexity while still keeping it simple to code and express.

 In this work we choose to start from a uniform density and so
apply no perturbation to this quantity. We take $\rho_0=1$,
$P_0=10^{-6}$, $\Omega=c_s=10^{-3}$ and $q$ is taken to be that for
a Keplerian profile namely, $3/2$. The embedded magnetic field is
uniform, in the vertical-direction and such that $\beta=800$.

Initially, unlike in most of the earlier works we choose to consider
a  shearing box where each side is of unit length i.e.\ one that has
the same length and number of grid points in each directions. Such
an approach facilitates probing to higher resolutions without
resorting to earlier methods of assuming \textit{apriori} that less
resolution is required in the y-direction. While less resolution
might be warranted in the y-direction in simulations that have a
clear difference in the scale of the structures in each of the
spacial directions we choose here to use caution and not make this
assumption which, amongst other things, gives rise to an anisotropy
in the numerical dissipation.

The equations, given above, are solved with a parallel
finite-difference time-explicit code. The code is essentially just a parallelized
version of the serial ZEUS code that is freely available. As in earlier papers (e.g.\
\cite{SN92}; \cite{HS}; \cite{HGB1}) we use the method of
constrained transport scheme for magnetic advancement so that the
$\nabla. \textbf{B}=0$ is enforced to the accuracy of the machine.
In this work we use the standard shearing-box boundary conditions.
Thus the domain is periodic in both the $y$ and $z$ directions and
shearing-periodic in $x$ (see \cite{HGB1} for further details).This
code contains an adaptive step size that is multiplied, as is
standard, by a safety factor. We have conducted several tests to
ensure that the step size does not give rise to numerical
instability. This check involved varying the safety factor in this
code to ensure that all prominent global features, trends and
saturation levels are fully captured by the integration, which could
be referred to as a check that we have the `true' or `converged' solution.

\begin{table}
 \centering
  \caption{Summary of all the cases presented in the paper}
  \begin{tabular}{@{}ccccc@{}}
  \hline
 Case & Resolution & Aspect & Artificial & Mean
 \\
  &  & Ratio& Viscosity &
 Cleaning\\

 \hline
1& $32:32:32$& 1:1:1& No & No\\
2& $64:64:64$& 1:1:1& No & No\\
3& $128:128:128$& 1:1:1& No & No\\
4& $32:32:32$& 1:1:1& Yes & No\\
5& $64:64:64$& 1:1:1& Yes & No\\
6& $32:32:32$& 1:1:1& No & Yes\\
7& $64:64:64$& 1:1:1& No & Yes\\
8& $128:128:128$& 1:1:1& No & Yes\\
9& $64:128:64$& 1:2:1& No & Yes\\
10& $64:192:64$& 1:3:1& No & Yes\\
11& $64:384:64$& 1:6:1& No & Yes\\
\hline
\end{tabular}\label{table1}
\end{table}

\section{Results}

Before beginning to discuss the finding it helpful to provide an
overview of all cases at this point. Table \ref{table1} shows each
of the cases that we have considered. We have used the method
detailed in \cite{HGB} to ensure that the fastest growing wavelength
is resolved.

We will choose to split the results section into two halves. Firstly
we will discuss the effect of increasing the resolution and then we
will go on to discuss the effect of increasing the box length, in
the $y$ direction, while maintaining the same resolution per unit
length.

\subsection{Ideal MHD}

We begin the discussion of the results with our fiducial model, case
1. In this model we consider a resolution of 32 grid points in each
of the three directions. This fiducial model is at a similar
resolution, in the $x$ and $y$ directions, as can be found in many
of the original calculations. This said, the earlier calculations did typically a
longer box length than we do here. This deliberate
change was made for two reasons. First, the decrease in the box size enables
us to push to much higher resolutions. Second, and most importantly
for what we wish to examine here, an increased integration time
i.e.\ such a reduction in the box length facilitates being able to
ascertain many more orbits for the same computational cost. We shall
discuss the ramifications of increasing the box length later in the
paper and some comments on the effect of varying the computational
domain can be found in \cite{WBH}.

\begin{figure}
\begin{center}
    \leavevmode\epsfxsize=8cm\epsfbox{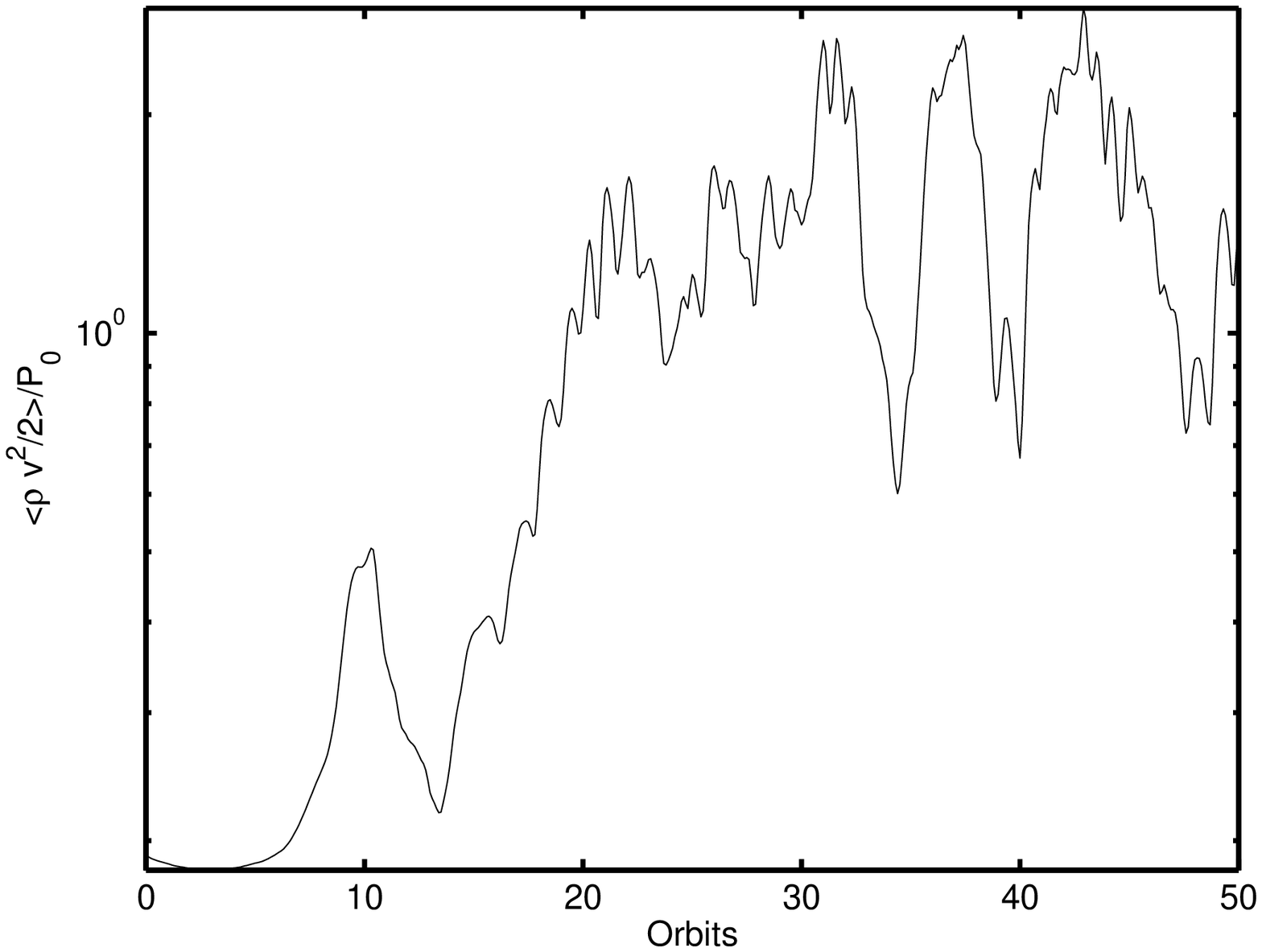}
    \leavevmode\epsfxsize=8cm\epsfbox{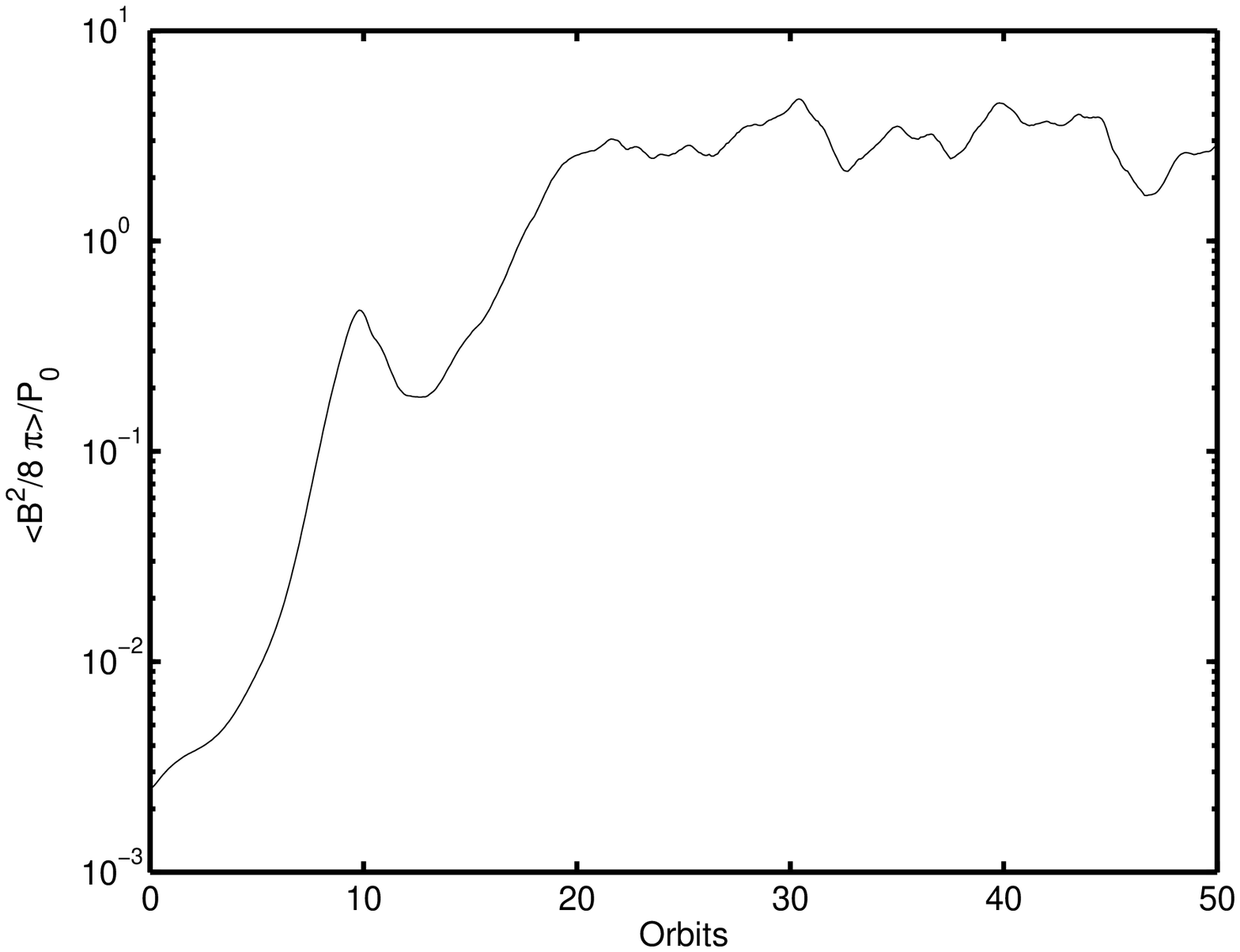}
    \caption{Top: Kinetic energy evolution for the fiducial case for the first 50 Orbits. Bottom: Magnetic energy evolution (normalised) for the fiducial
    case for the first 50 orbits.
}\label{fiden1}
\end{center}
\end{figure}

As stated in the introduction, we wish to determine a cumulative
average for the quantity $\alpha$, which is related to the
difference of the Reynolds and Maxwell stresses. Such a
quantity can only be calculated if the simulations are run for
sufficiently long so that there is a large time in the saturated
state for the cumulative average to settle. Figure \ref{fiden1} shows
the initial section evolution of the magnetic and kinetic energies
for the $32^3$ case and illustrates the it can take several orbits
for the saturated state to begin. We note that, despite the
reduction in the box size, these figures resemble those from the
earlier calculations. However, we note that time to reach a
statistically steady state occurs at a later point in terms of the
number of orbits but at a comparable time from the start of the
simulations. Other slight differences from earlier calculations are
combined effect of box size, isothermality and choice of plasma $\beta$.

\begin{figure}
\begin{center}
    \leavevmode\epsfxsize=8cm\epsfbox{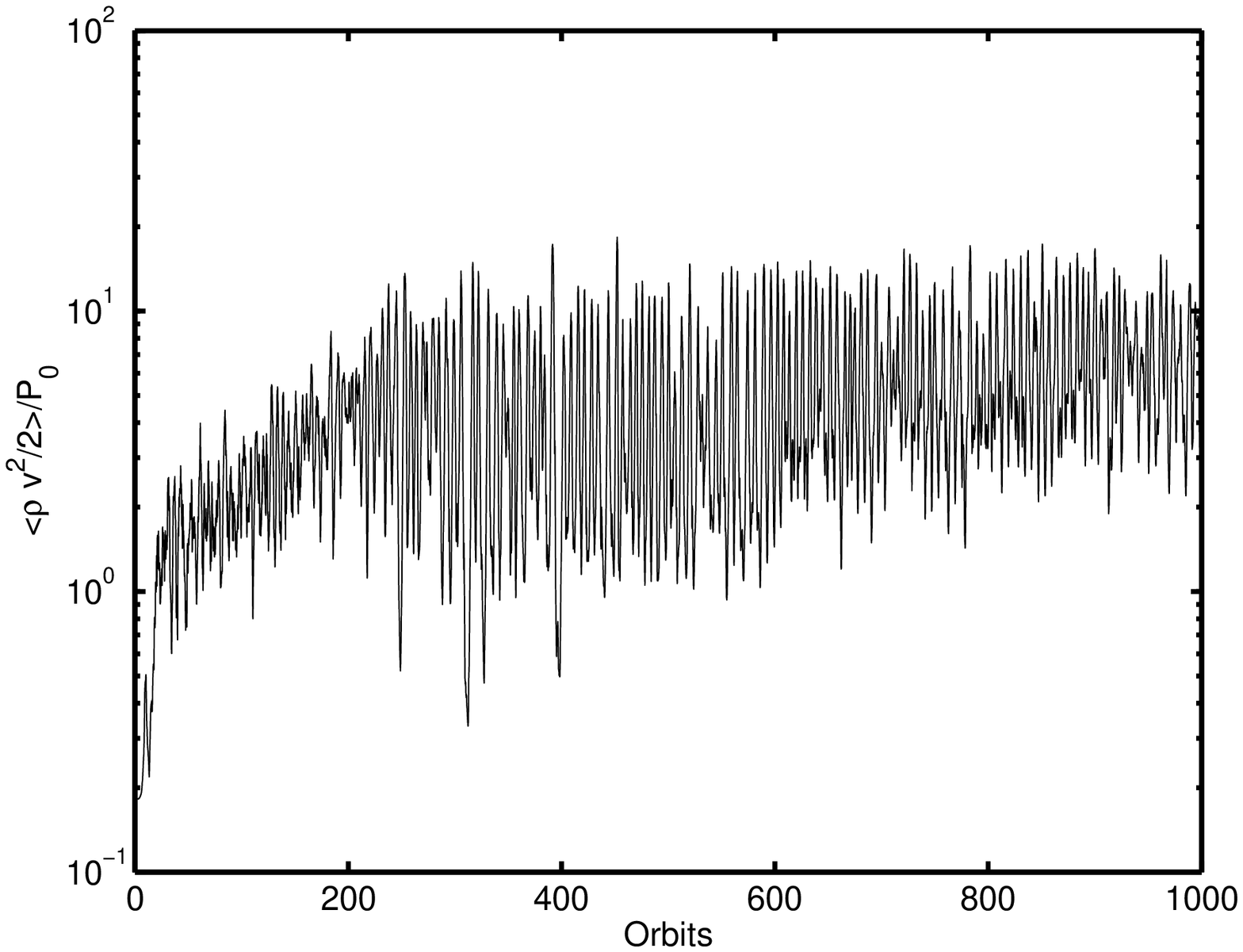}
    \leavevmode\epsfxsize=8cm\epsfbox{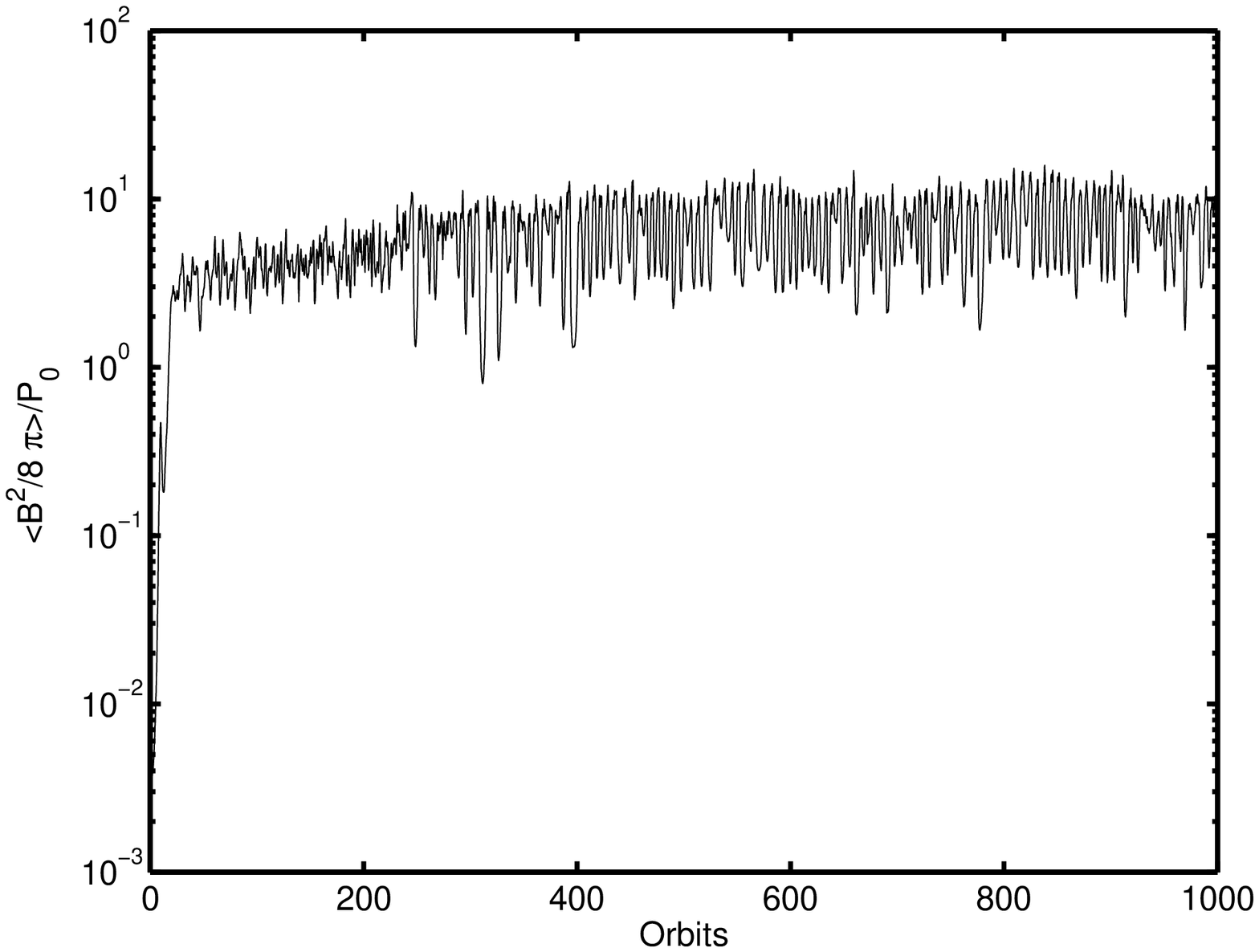}
    \leavevmode\epsfxsize=8cm\epsfbox{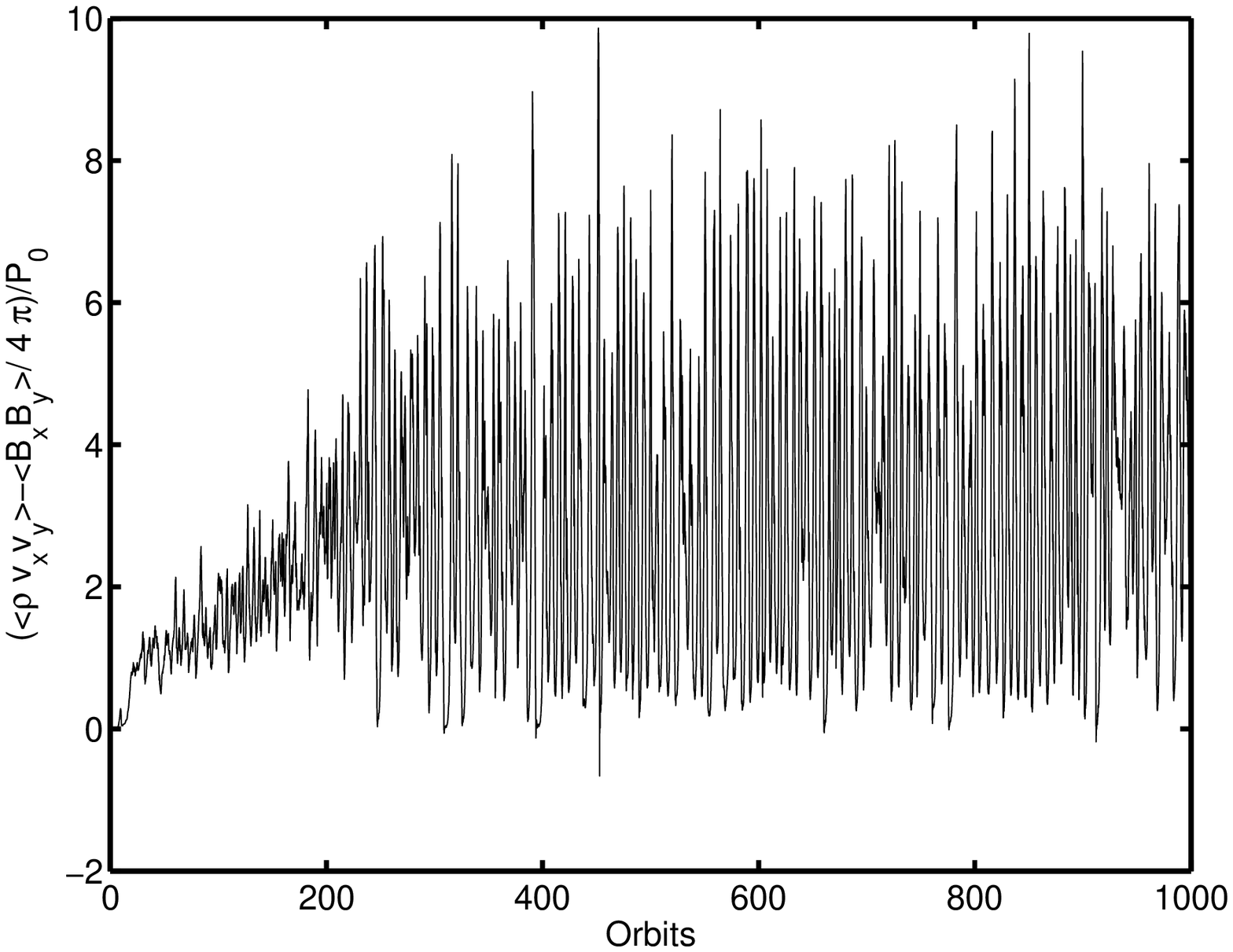}
    \caption{Top: Kinetic energy evolution (normalised) for the
      case 1 for 1000 orbits. Middle: Magnetic energy evolution
      (normalised) for case 1 for 1000 orbits. Bottom: The evolution of $\alpha$ for case 1.
}\label{fiden2}
\end{center}
\end{figure}

These figures appear to show
that the evolution is complete and that the final state has been
obtained and thus one might thing that all that remains  to
calculate a cumulative average is to carry the calculation on for
sufficient time and then evaluate the cumulative average of
$\alpha$.
However, figure \ref{fiden2} shows that, in this calculation, what
appears to initially be a statistically steady state is only
approximately so. If you examine the solution for 1000 orbits one
sees that the first 200 orbits or so is in fact a slow ramp up to a
higher saturation level.

\begin{figure*}
\begin{center}
    \leavevmode\epsfxsize=6cm\epsfbox{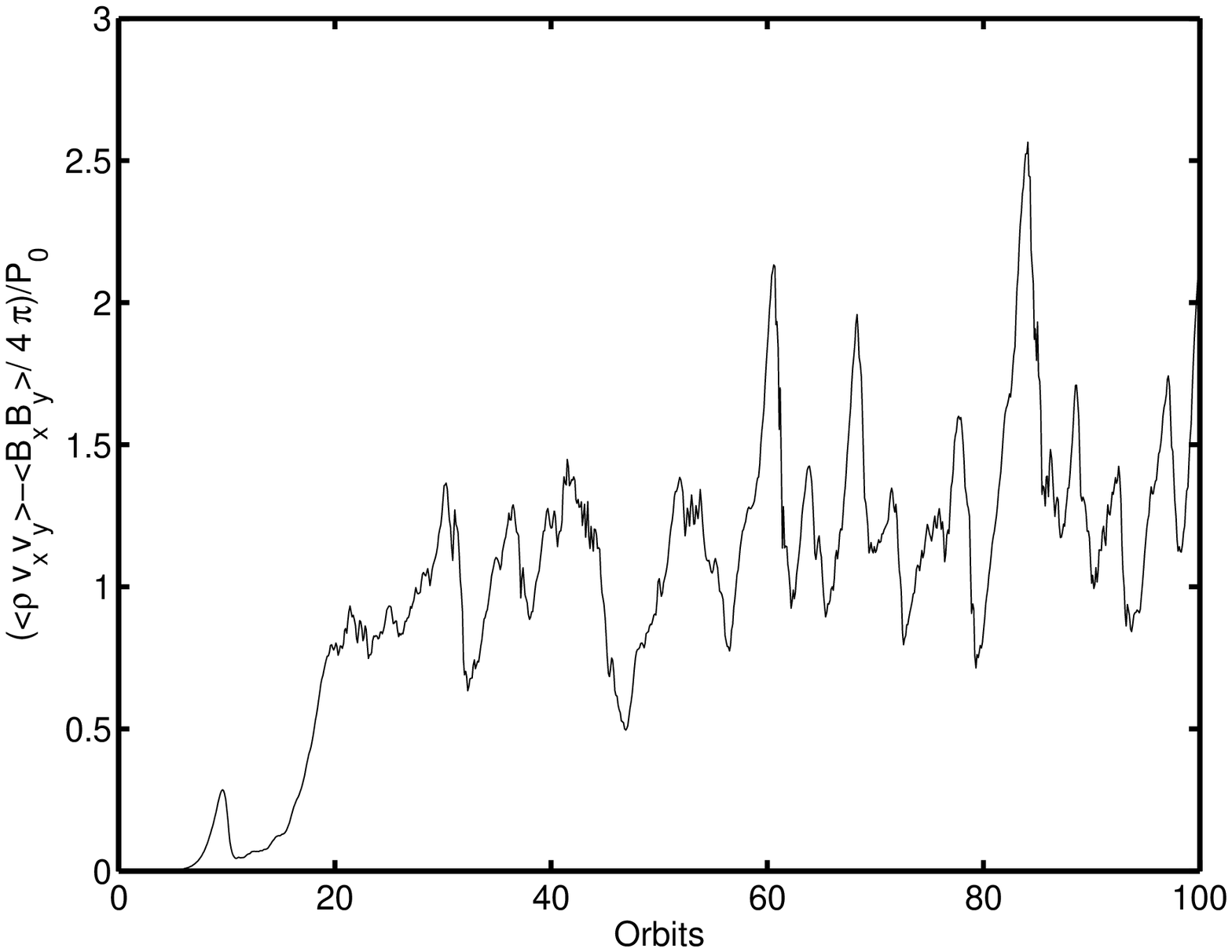}
    \leavevmode\epsfxsize=6cm\epsfbox{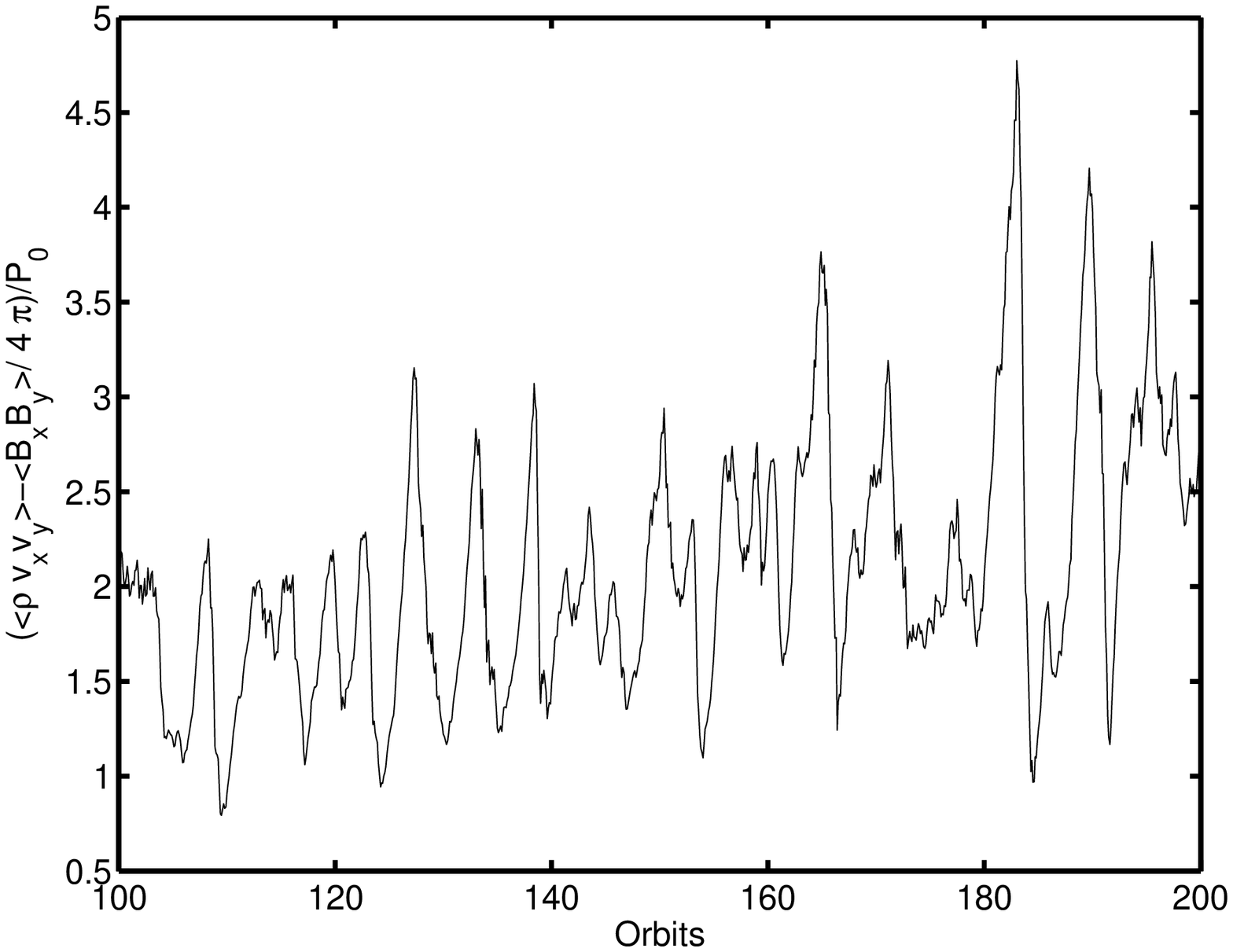}
    \leavevmode\epsfxsize=6cm\epsfbox{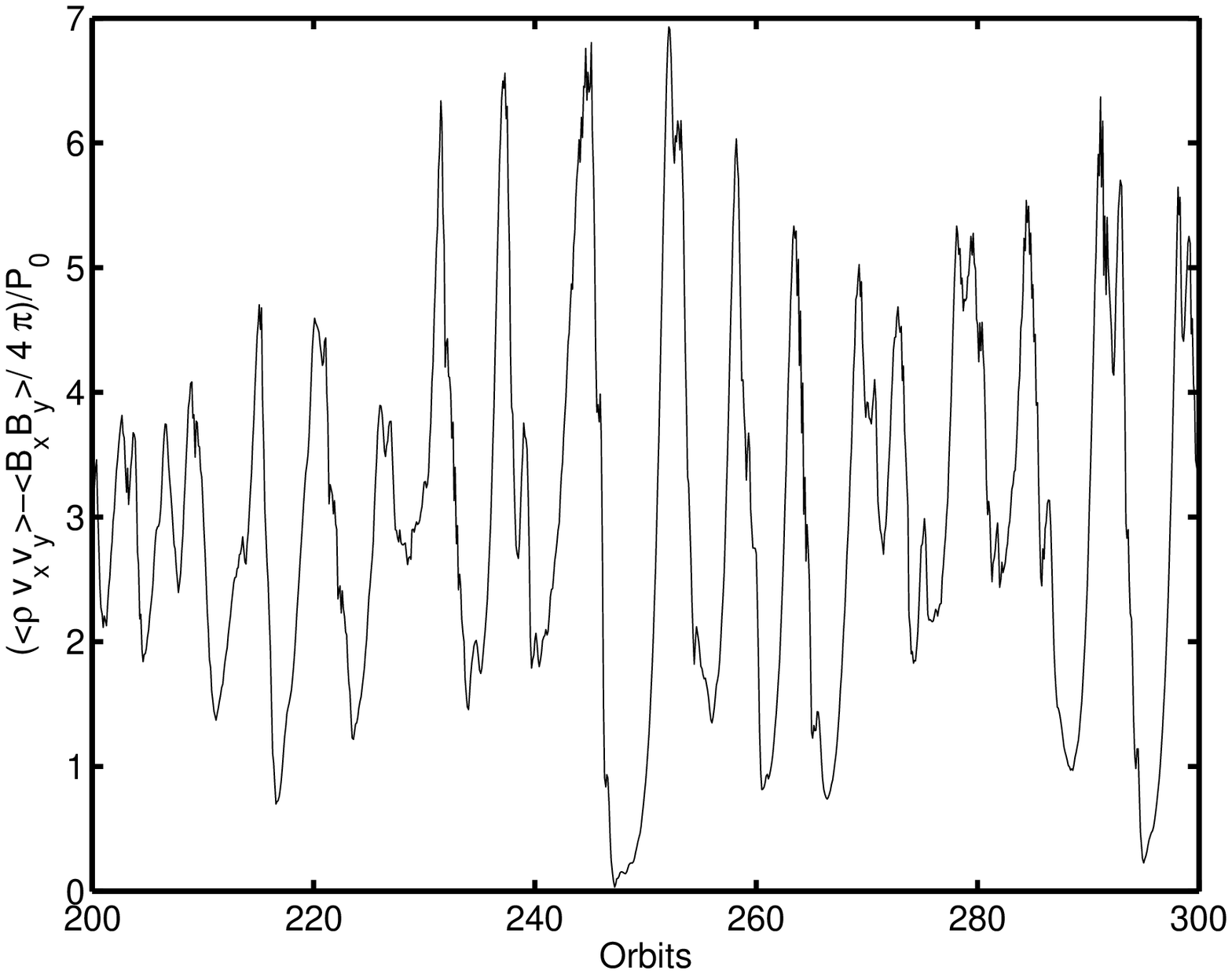}
    \leavevmode\epsfxsize=6cm\epsfbox{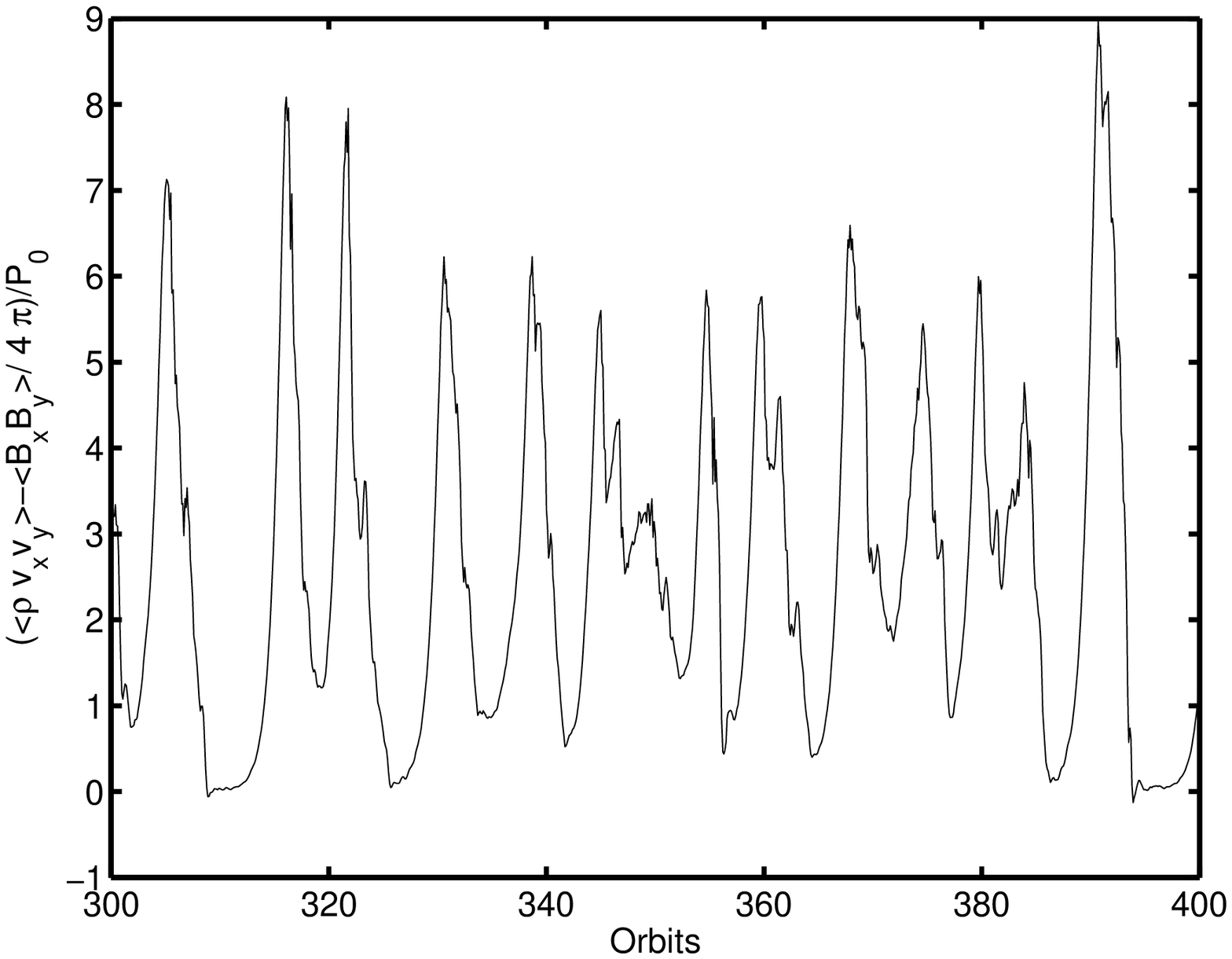}
    \leavevmode\epsfxsize=6cm\epsfbox{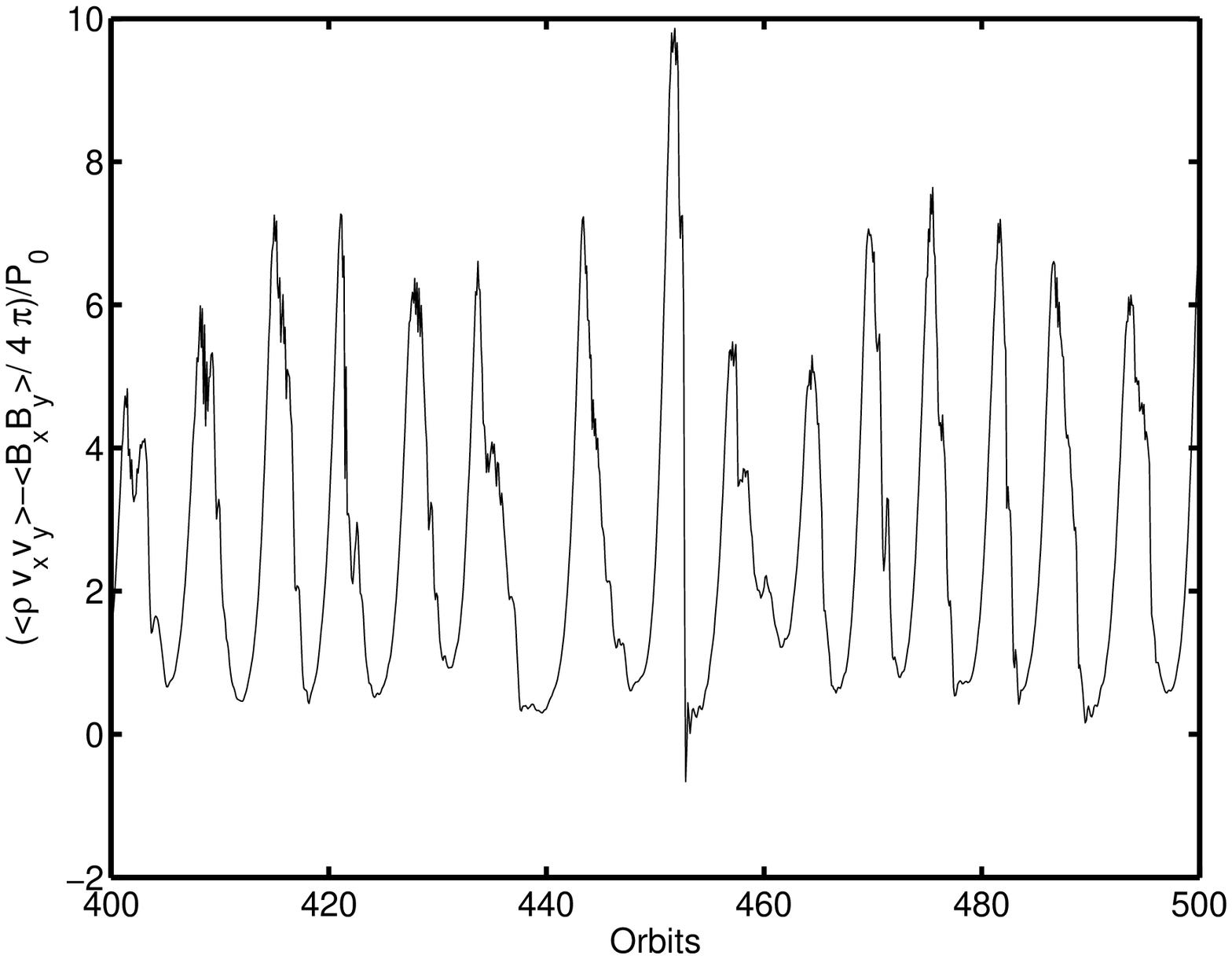}
    \leavevmode\epsfxsize=6cm\epsfbox{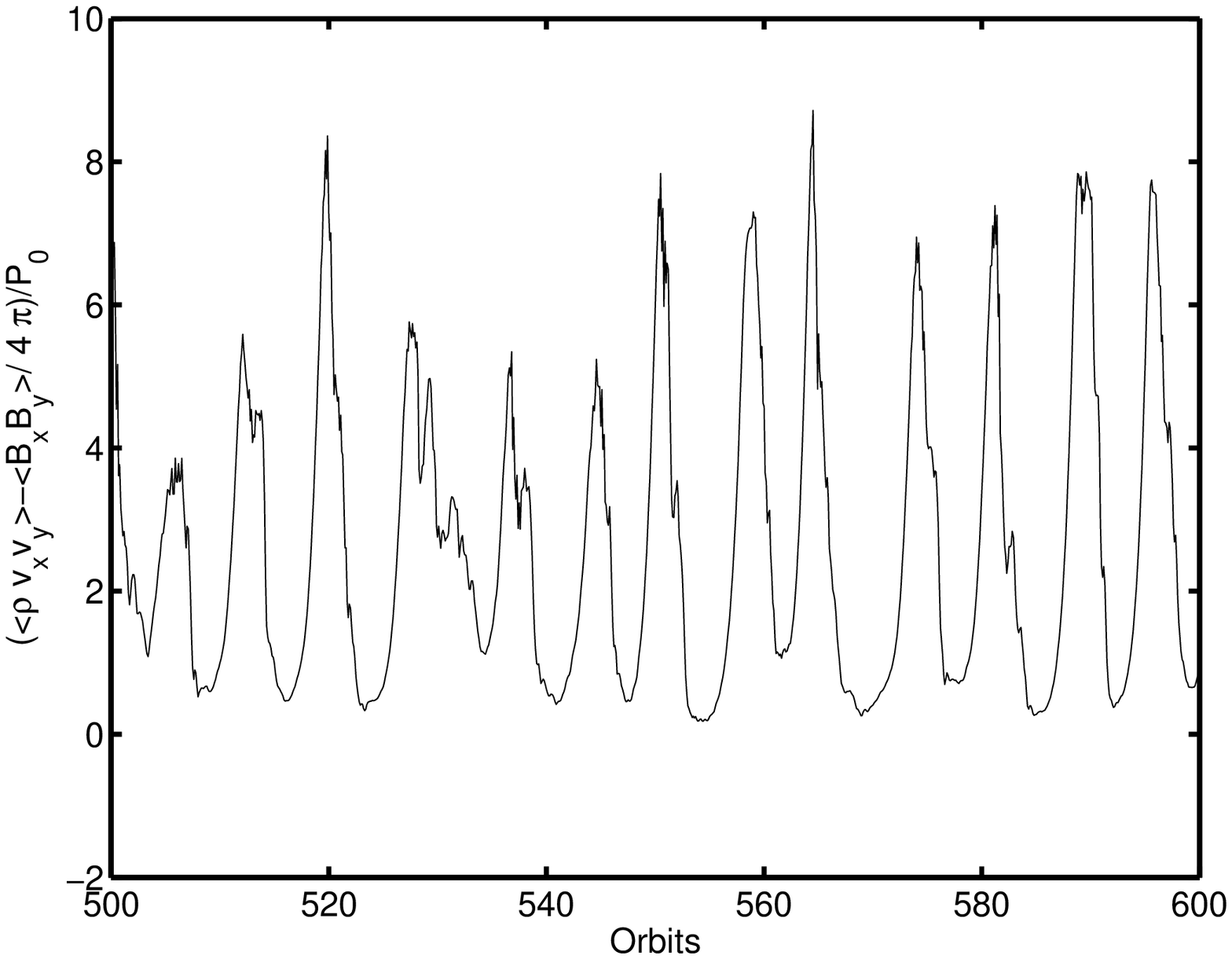}
    \leavevmode\epsfxsize=6cm\epsfbox{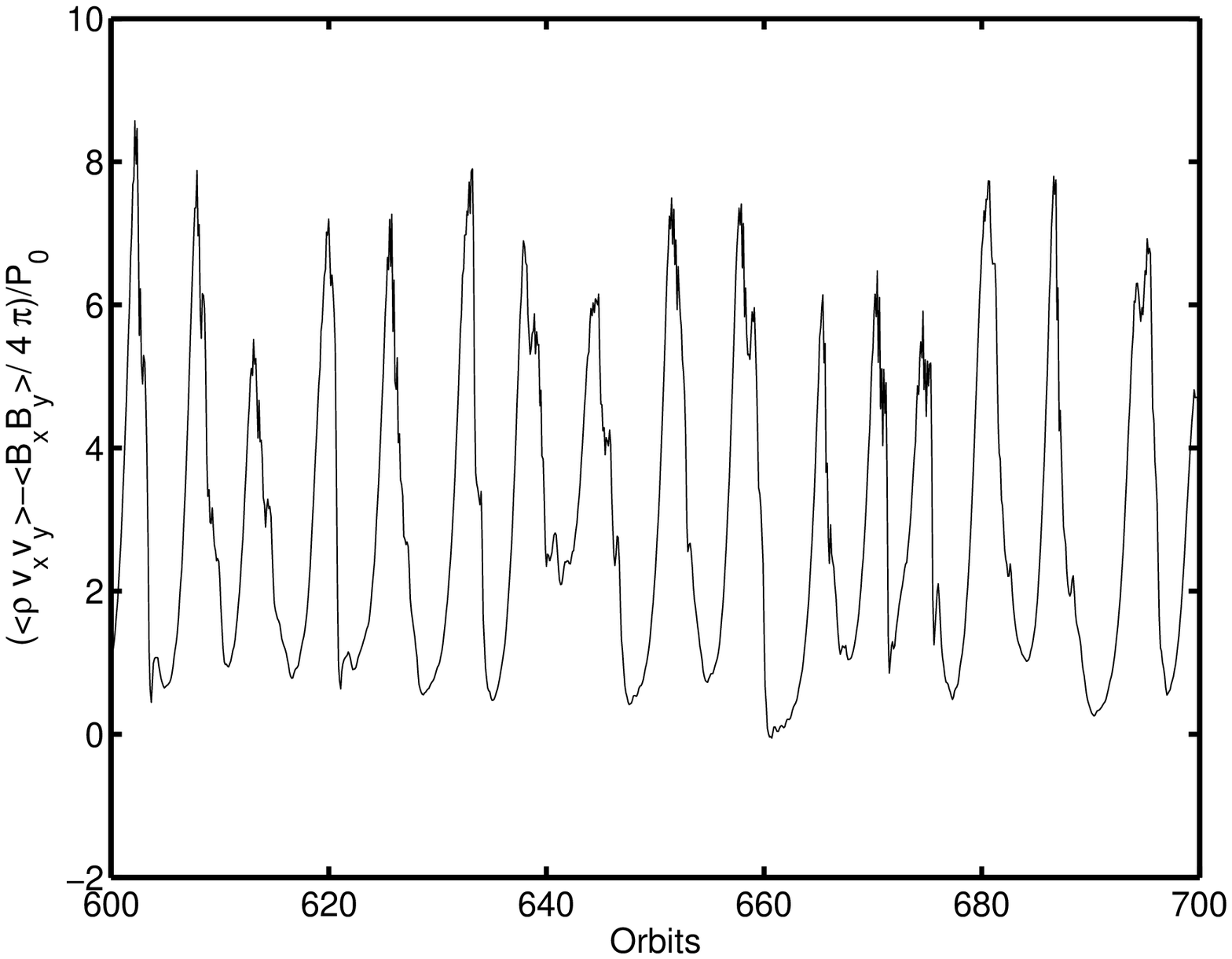}
    \leavevmode\epsfxsize=6cm\epsfbox{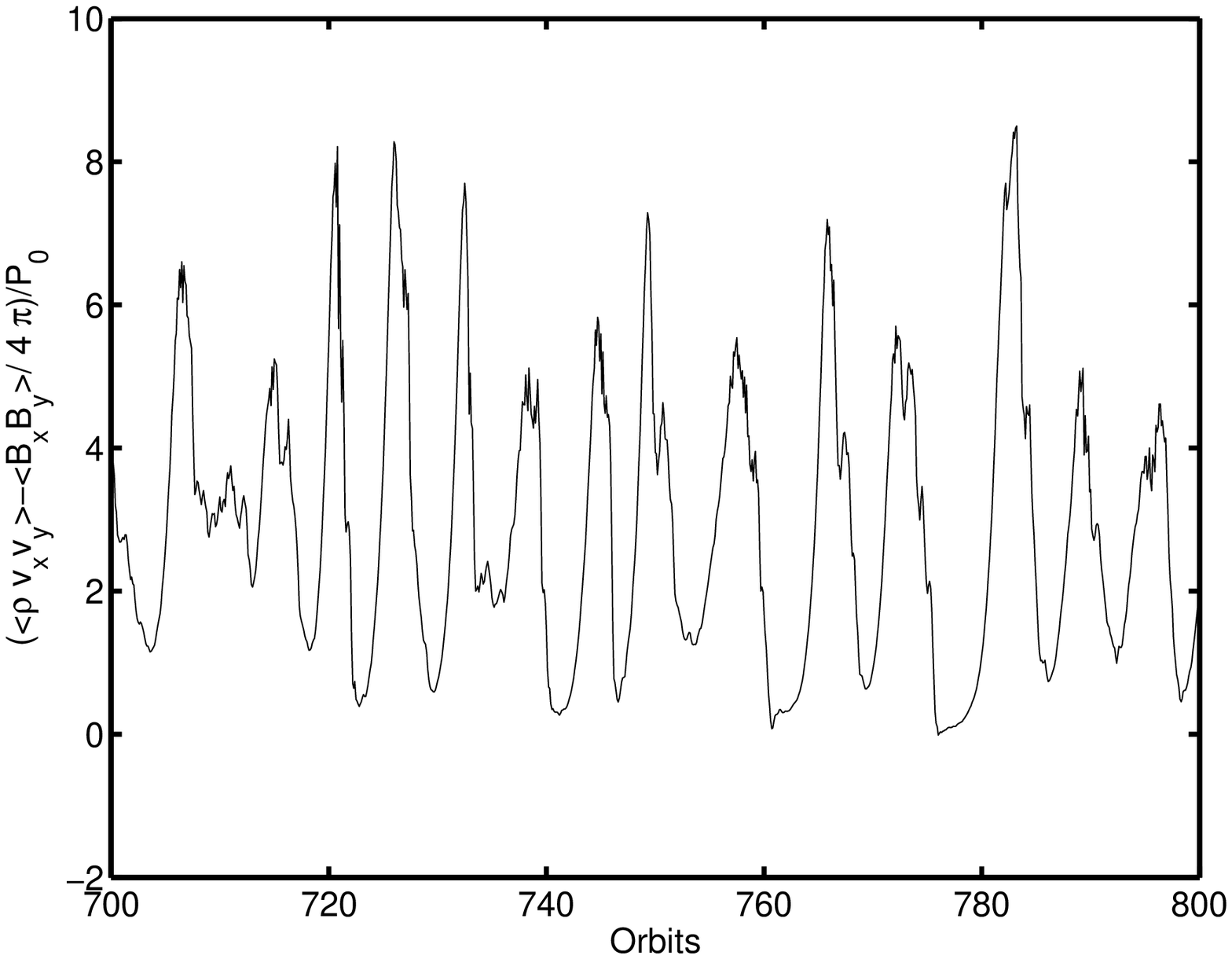}
    \leavevmode\epsfxsize=6cm\epsfbox{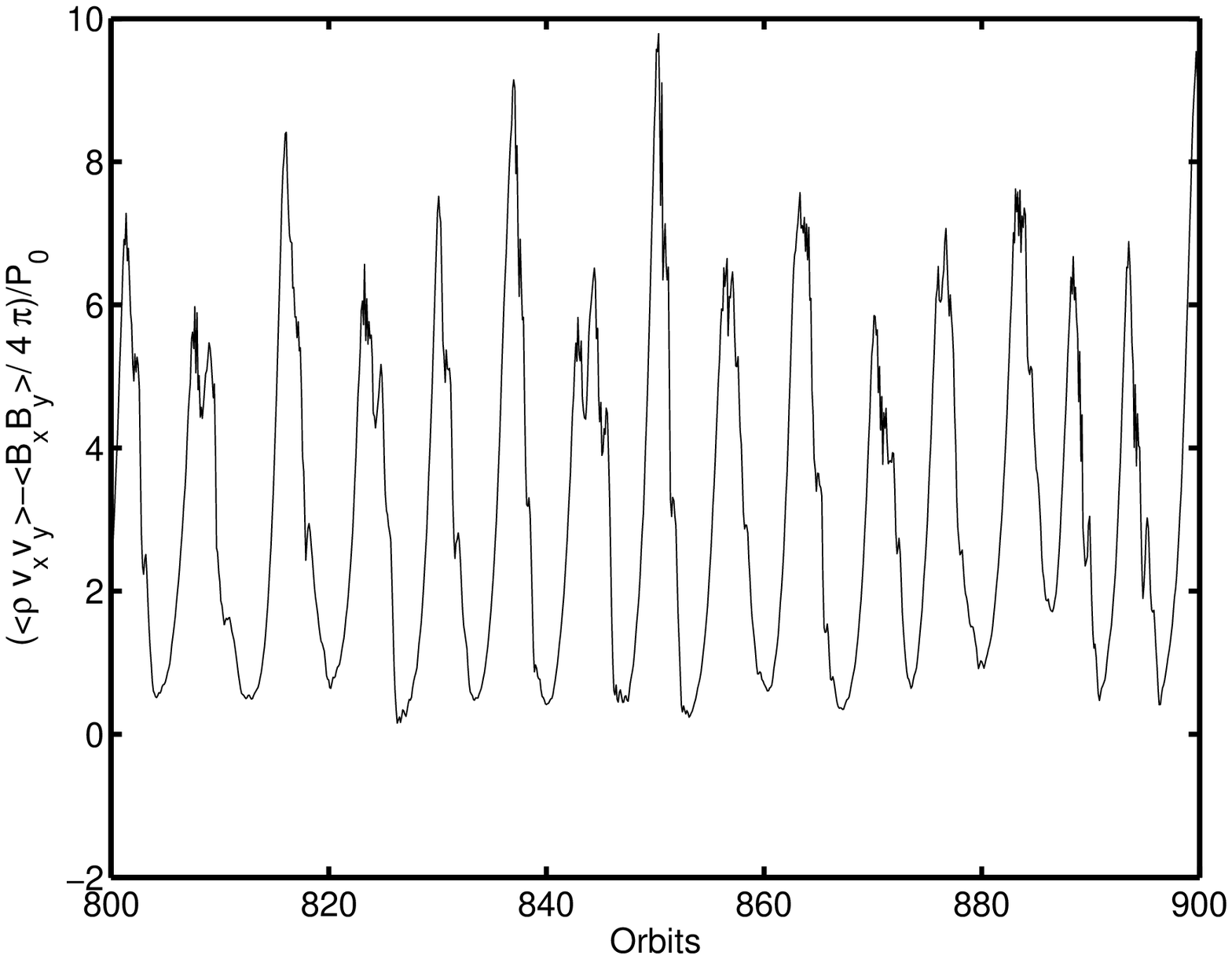}
   \leavevmode\epsfxsize=6cm\epsfbox{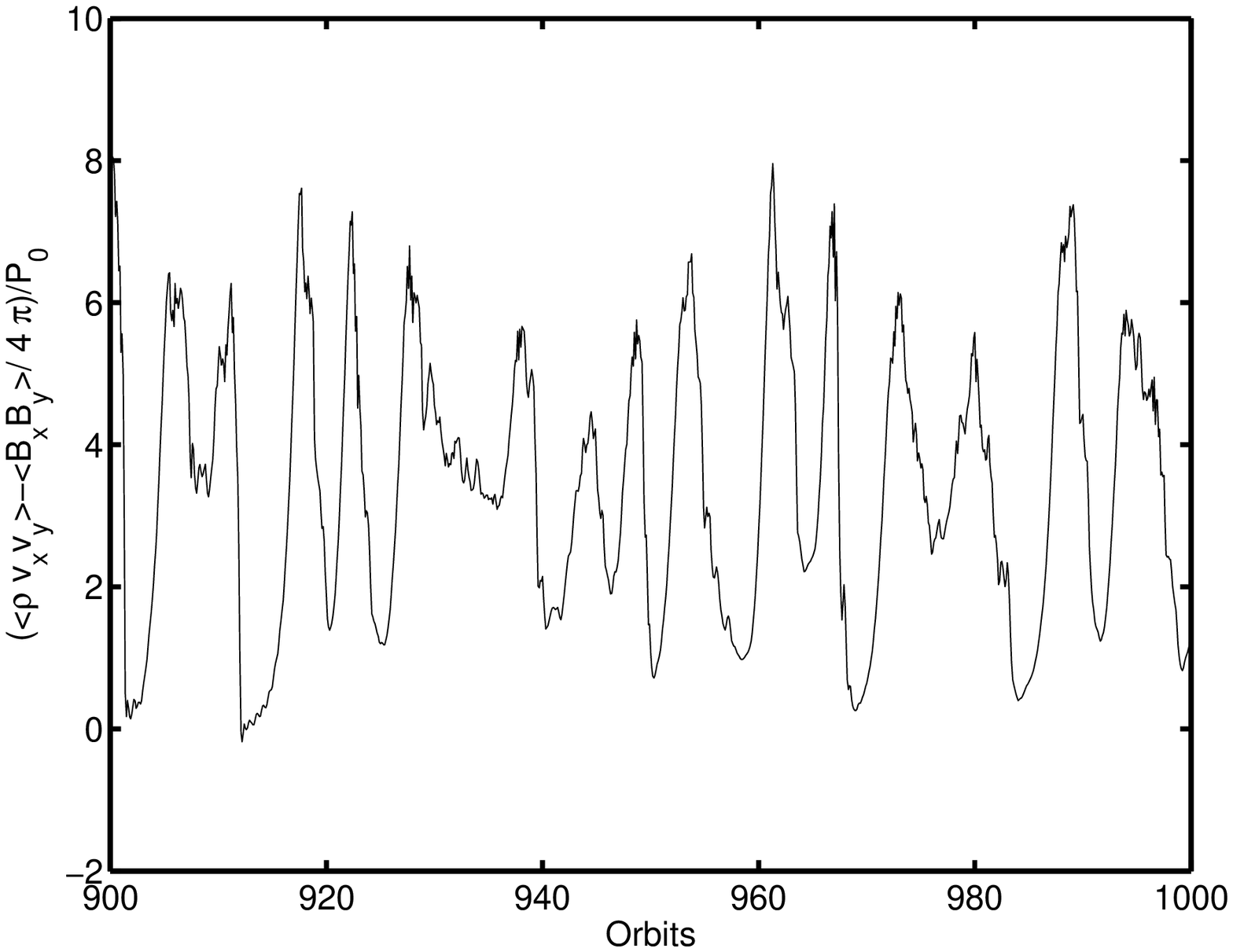}

    \caption{A series of plots showing in more detail the evolution of
    $\alpha$ as the number of Orbits increases for case 1}\label{cumulative32}
\end{center}
\end{figure*}

The most important implication from a shift in the average magnetic
and kinetic energies is that this is almost certainly a
corresponding shift in the momentum transport and figure \ref{fiden2}
shows that this is indeed the case. However, from a figure of this
kind it is exceptionally hard to gain a feeling for the structure
---indeed much of the information on the temporal variability is
being lost. As such, we provide a more detailed series of snapshots
over the full 1000 orbits in \ref{cumulative32}. This images show that
the system is moving from a state which is rather turbulent into a
state where there is a repeated classical channel solution
presenting before being destroyed and the cycle repeating.

\begin{figure}
\begin{center}
    \leavevmode\epsfxsize=8cm\epsfbox{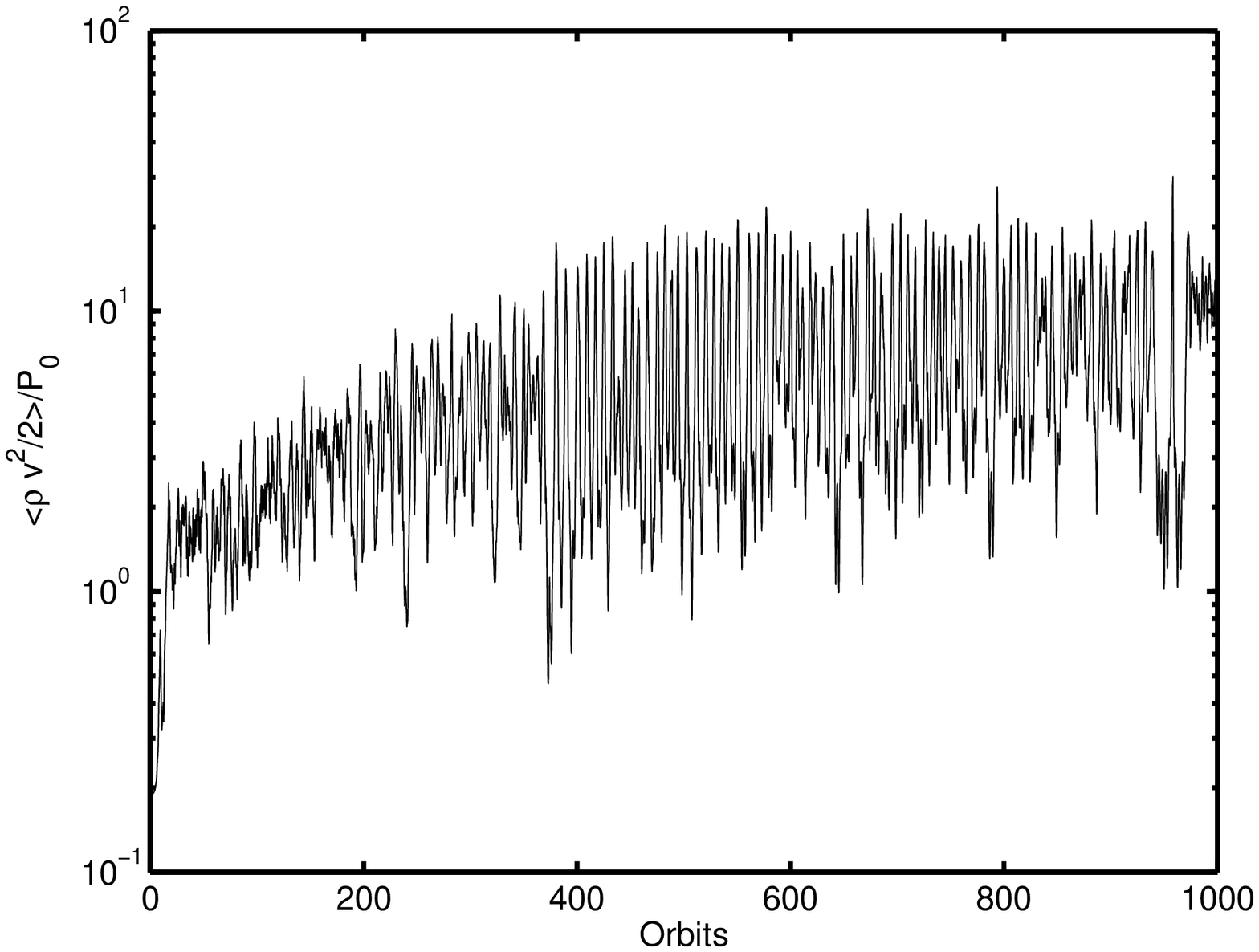}
      \leavevmode\epsfxsize=8cm\epsfbox{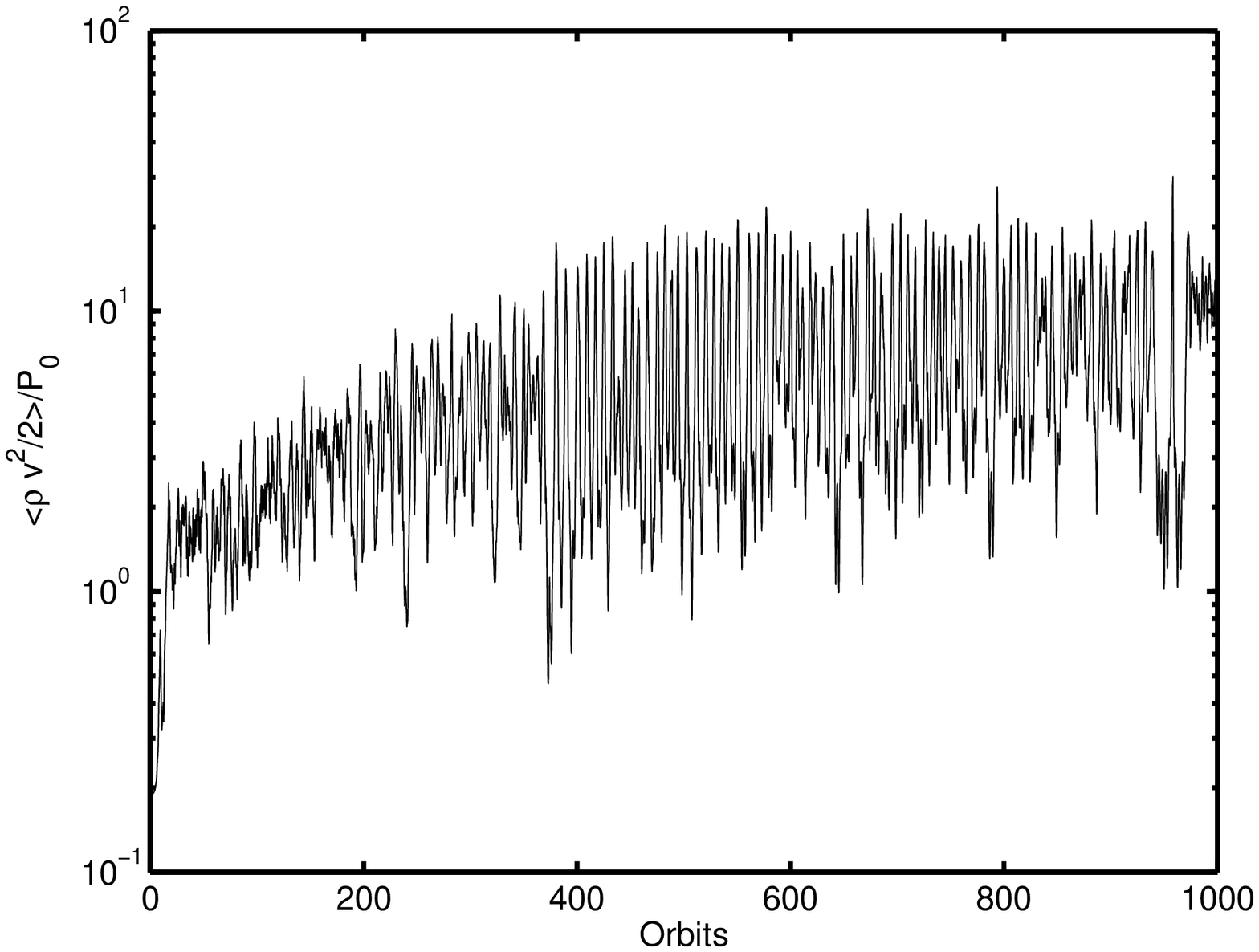}
    \leavevmode\epsfxsize=8cm\epsfbox{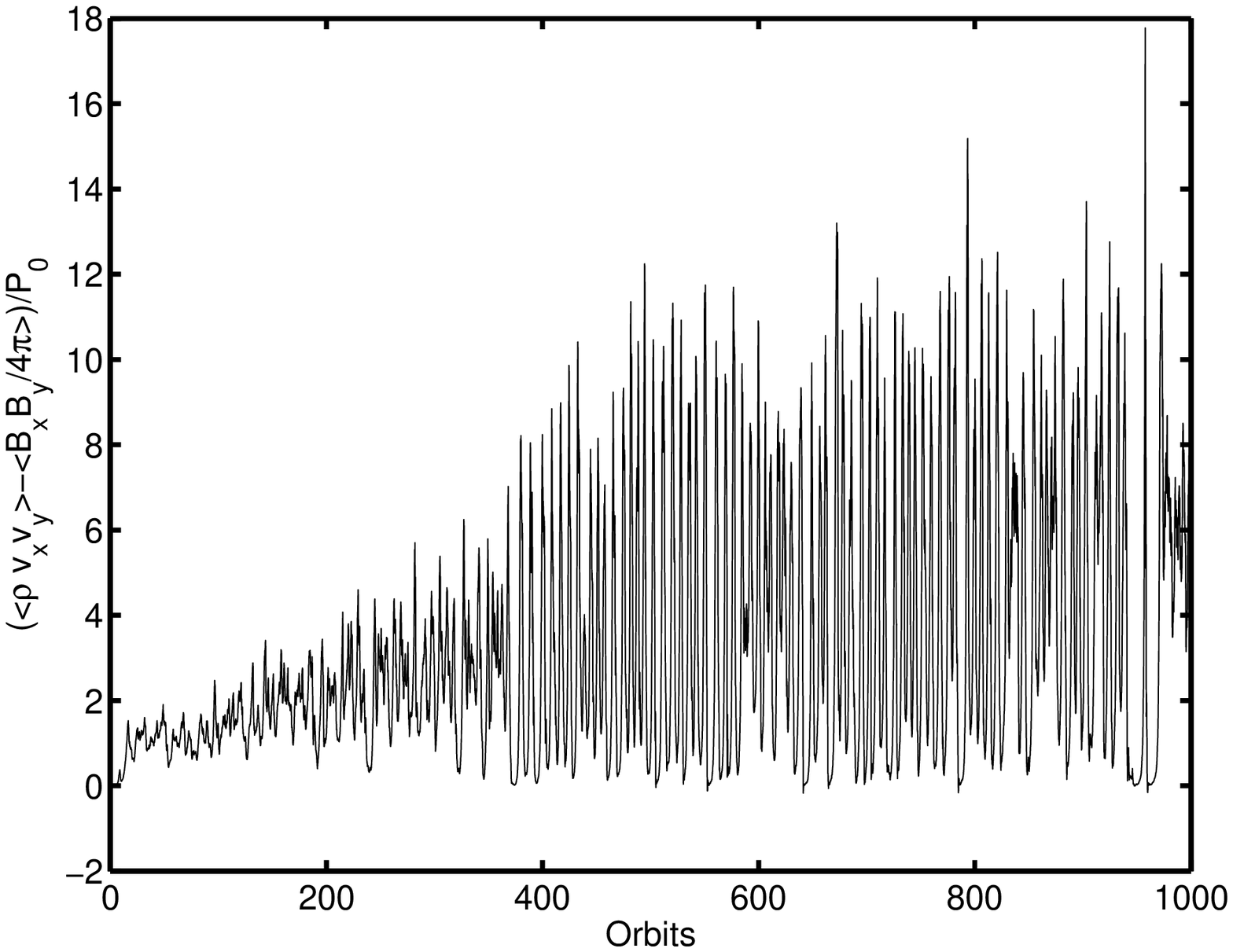}
\caption{Top: Kinetic energy evolution (normalised) for case 2 for
1000 orbits. Middle: Magnetic energy evolution (normalised) for case
2 for 1000 orbits. Bottom:The evolution of $\alpha$ for the case where there are $64^3$ grid points (case 2).}\label{kin_64}
\end{center}
\end{figure}

In the light of the images show above, one of the most natural
questions concerns whether the same behaviour is still found if the
length-scale for the dissipation scale is decreased. In this problem
this is akin to increasing the number of grid points that are used.
As such we have examined the case when the number of grid points in
each spatial direction and the findings are shown in figure
\ref{kin_64}. Note that, while the climb to the
saturated state is slightly slower in this case, a transition from
one quasi-statistically steady state to another still occurs.
Similar was found at a resolution of $128^3$ and so we conclude that
this behaviour is independent of the length-scale for dissipation.

\subsubsection{Artificial Viscosity}

In the above calculations we have not included a resistivity or a
viscosity of any kind and have just left the dissipation to be the
result of the grid. However, there in several of the local shearing
box calculations the was an artificial viscosity term included in
the equation of motion of the form (\cite{HGB1}):

\begin{equation}
A=C \rho (\delta_i v_i)²(\Delta x)
\end{equation}

The motivation for such a term was to provide limited extra
dissipation in the case when there is a strong compressional
solution. As such, it retains the original ideal magnetohydrodynamic
(MHD) nature of the equations and only calls for extra assistance is
special situations. While we would argue that a viscosity of some
kind is useful we would argue that it is much better to consider a
physical viscosity that acts continually.

However, while we do not believe that the artificial viscosity
method is the most instructive it is a method that has served the
community for many years and, as such, one must entertain the
possibility that the solutions found about might in some way be the
result of the lack of the inclusion of an artificial viscosity. As
such, for completeness we here examine the $32^{3}$ calculation with
the artificial viscosity term included and where $C=2.0$ as is
standard. This implies that smoothing is carried out over two cells.
Further discussion of the artificial viscosity can be found in
\cite{HGB1} and \cite{SN92}.

\begin{figure}
\begin{center}
    \leavevmode\epsfxsize=8cm\epsfbox{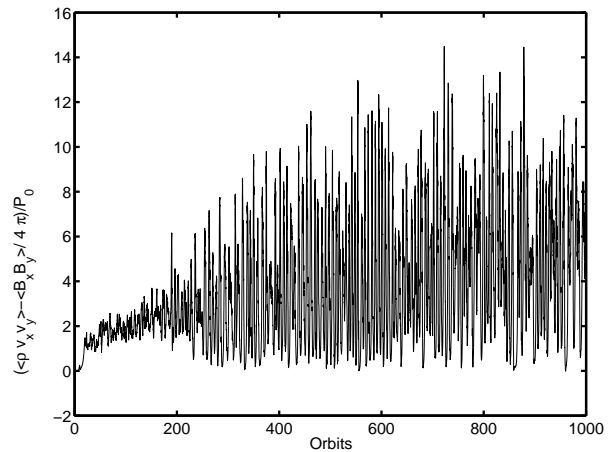}
    \caption{The evolution of $\alpha$ for the case where there are $32^3$ grid points and an artificial viscosity (case 4).\label{alphafullav}}
\end{center}
\end{figure}

Figure \ref{alphafullav} shows that the same global trend is
witnessed in this case as in the case where there is no artificial
viscosity included. This shows that the inclusion of an artificial
viscosity does not change the result found above.

\subsubsection{Problems with the Shearing Boundary Conditions}

\begin{figure}
\begin{center}
    \leavevmode\epsfxsize=8cm\epsfbox{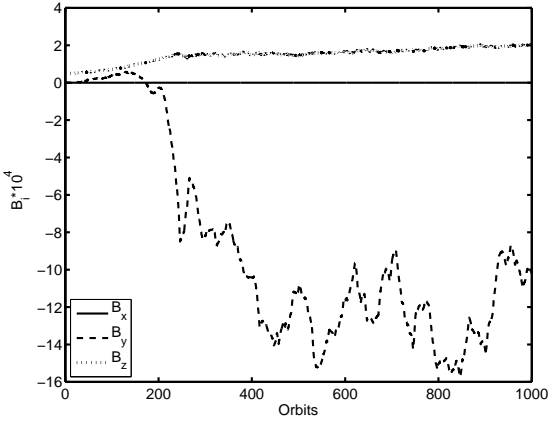}
\caption{ The evolution of each of the components of the
    magnetic field for case 1.}
\label{Bcomps}
\end{center}
\end{figure}

The behaviour witnessed above is fully explained by an examination
of what occurs to the mean of the various components. Many of these
components should formally be zero but the interpolation necessary
in the implementation of the shearing boundary conditions give rise
to errors as shown in figure \ref{Bcomps}. These errors have been
mentioned before in earlier calculations \cite{HGB}. However, many
of the early calculations were for short time spans and there the
error was stated to be about $10^{-3}c_s$. while we confirm  that
this is true for small numbers of orbits this is not true in the
long term. In fact, as figure \ref{Bcomps} shows the mean $B_z$ and
$B_y$ fields both grow and that the By field grows to such and
extent that it becomes of greater magnitude. We note here that the
numerical method employed prevents $B_x$ from obtaining significant
error. The constrained transport forces $B_x$ to stay constant
to round off because it is evolved with the emf components $E_y$ and
$E_z$ which are strictly periodic and hence $B_x$ is conserved to
round off error. However, both $B_y$ and $B_z$ have some error due
to the shearing box interpolation in $E_x$.

If we are to obtain a valid long-term average for $\alpha$ then we
need to remove this error, which compounds to give us an
unintentional switch in the mean field. As such we proceed by
forcing all mean values to their mathematically formal values by
removing the error at set intervals. Obviously the most rigorous
correction would be invoke this cleaning procedure every time-step.
However, this is computationally demanding. Therefore, we choose
here to correct the mean values every 100 time-steps as we
determined that this gives phenomenally good agreement with cleaning
every time-step for considerably less cost.

\begin{figure}
\begin{center}
    \leavevmode\epsfxsize=8cm\epsfbox{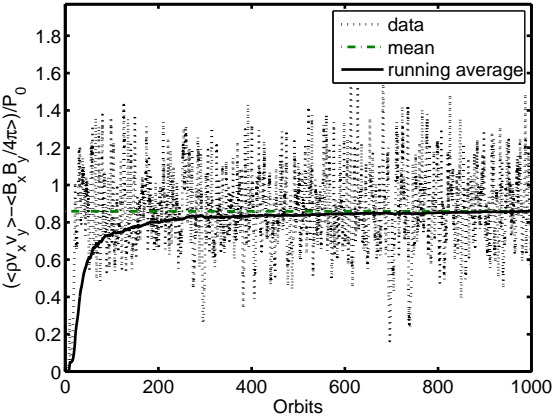}
    \caption{The evolution of $\alpha$ when full cleaning is employed
    for case 6 (1:1:1 aspect ratio)}
\label{remove1}
\end{center}
\end{figure}

\begin{figure*}
\begin{center}
    \leavevmode\epsfxsize=6.0cm\epsfbox{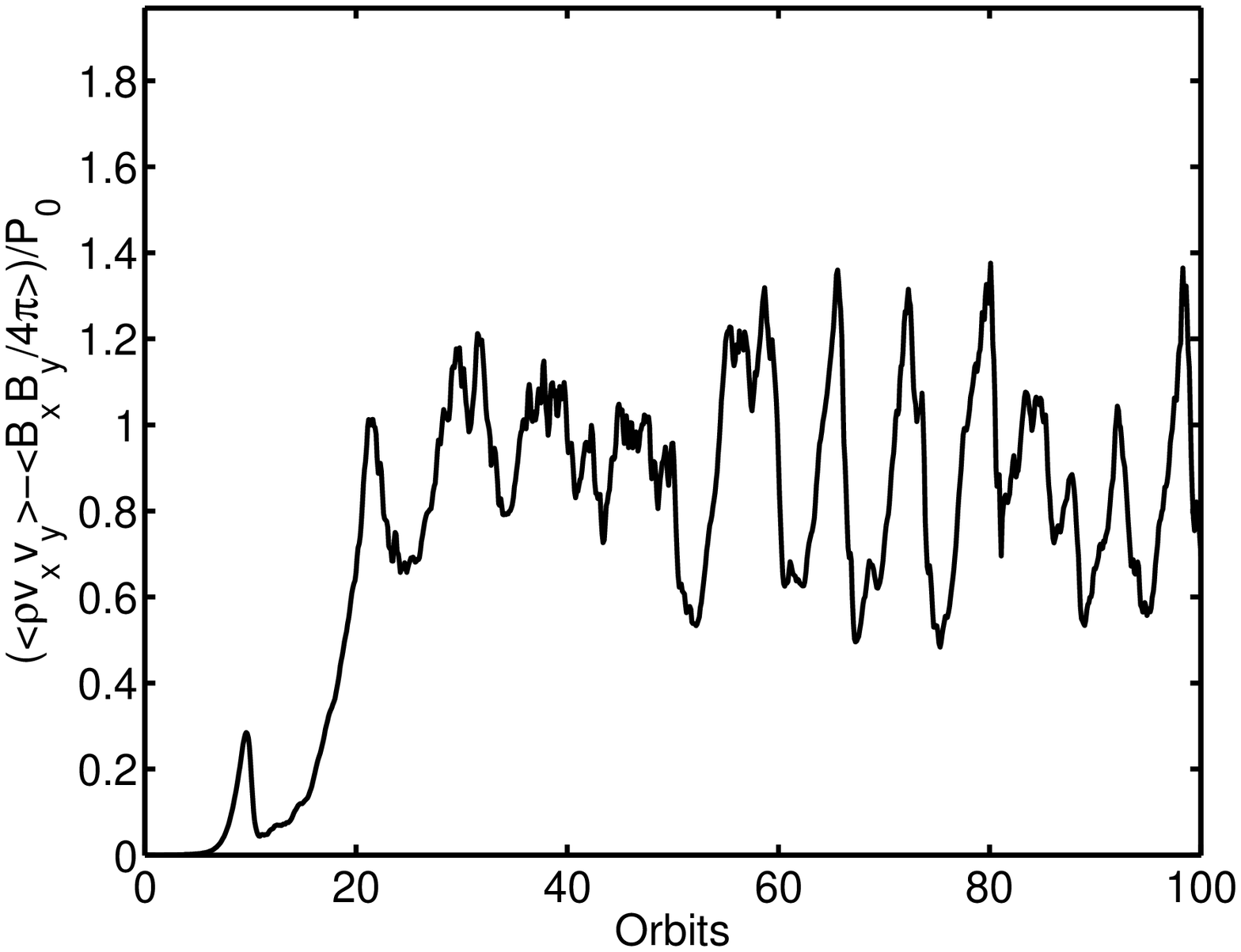}
  \leavevmode\epsfxsize=6.0cm\epsfbox{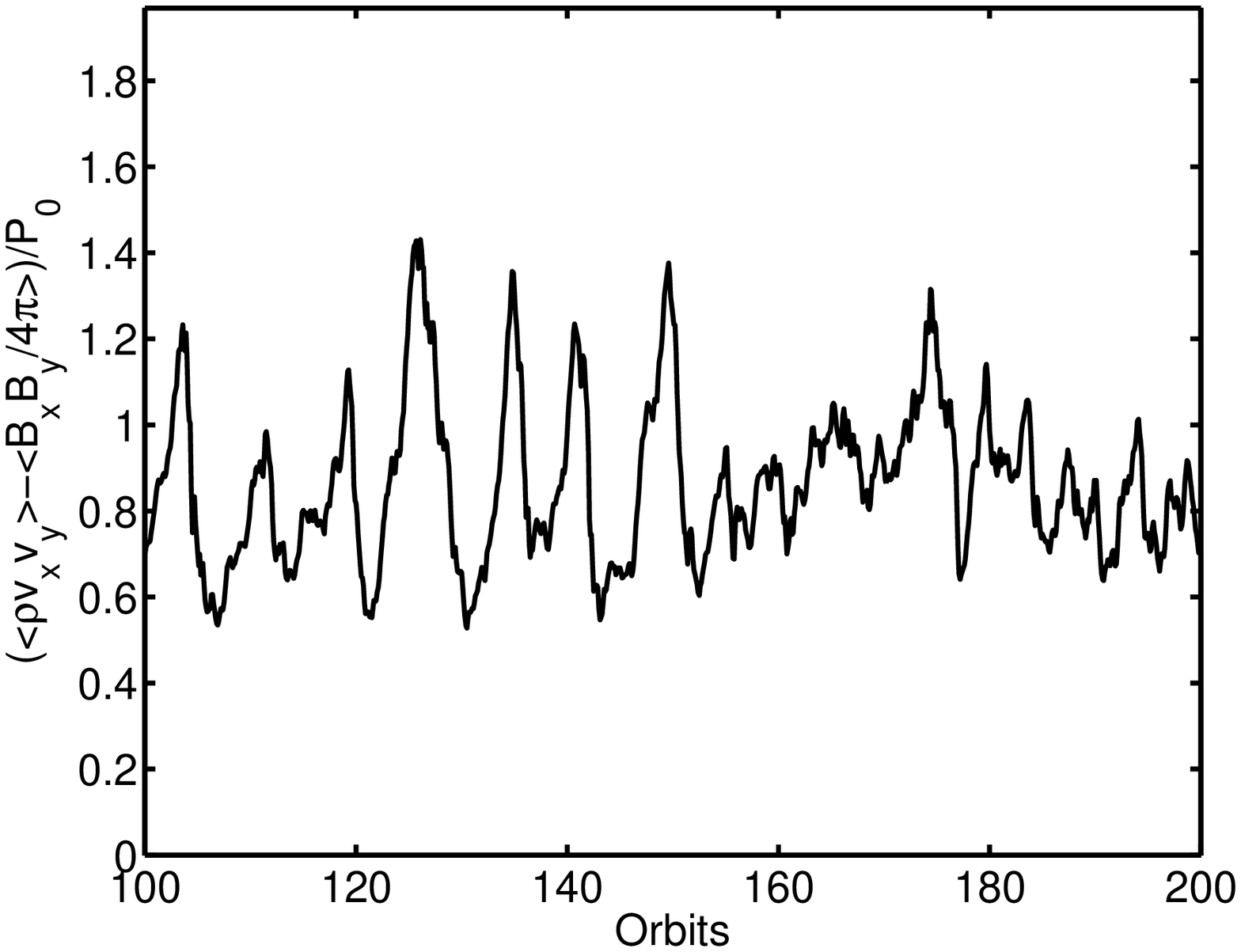}
  \leavevmode\epsfxsize=6.0cm\epsfbox{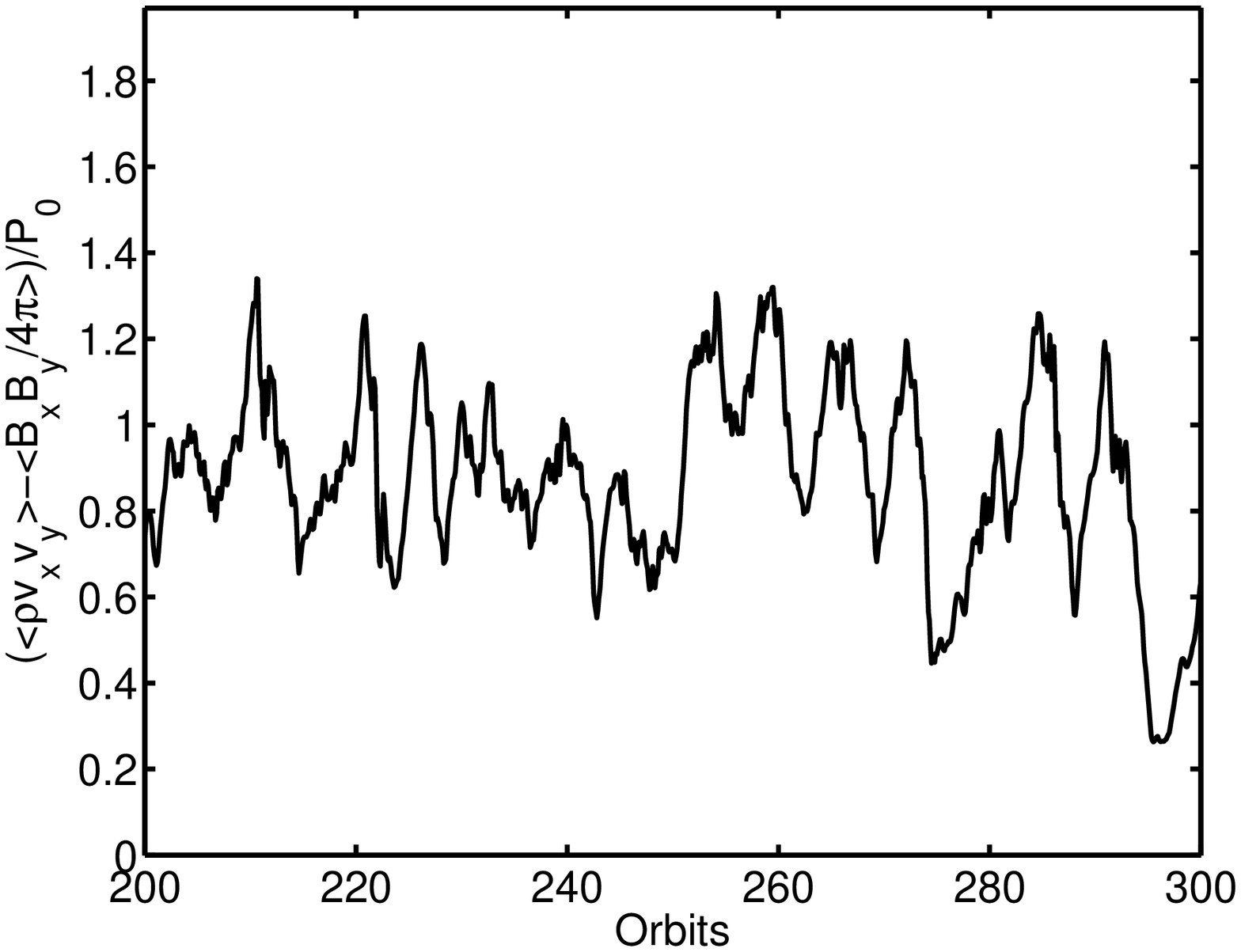}
  \leavevmode\epsfxsize=6.0cm\epsfbox{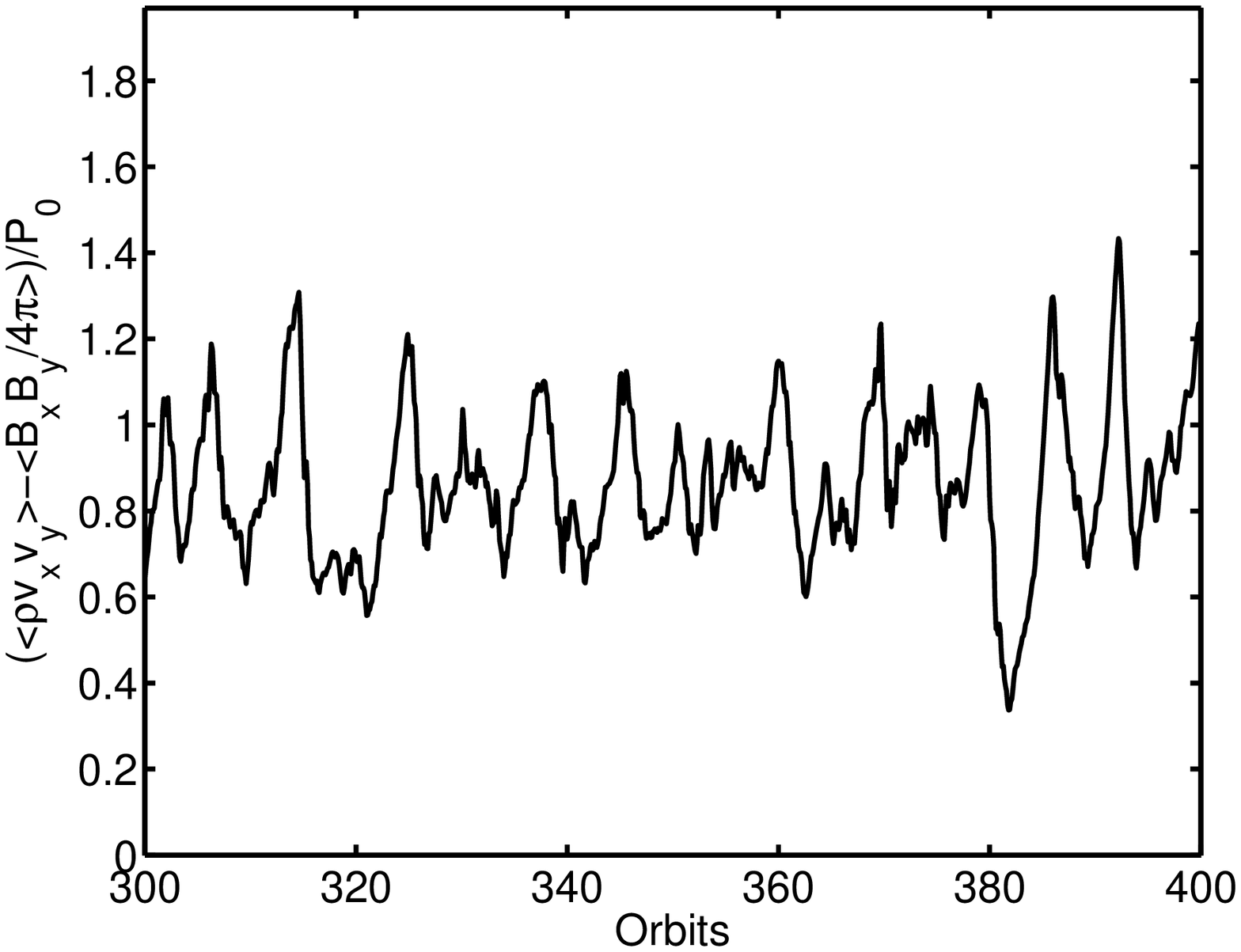}
  \leavevmode\epsfxsize=6.0cm\epsfbox{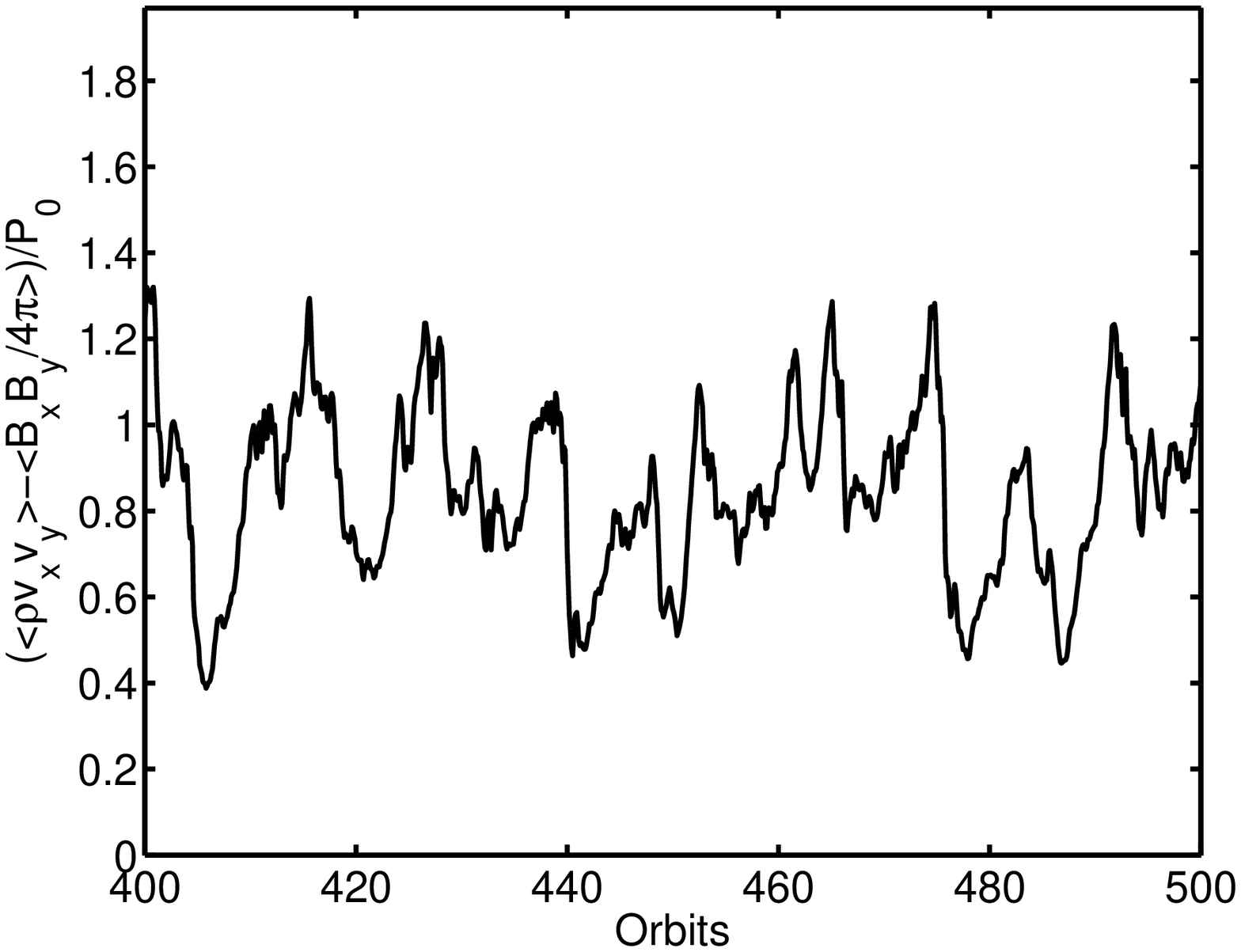}
  \leavevmode\epsfxsize=6.0cm\epsfbox{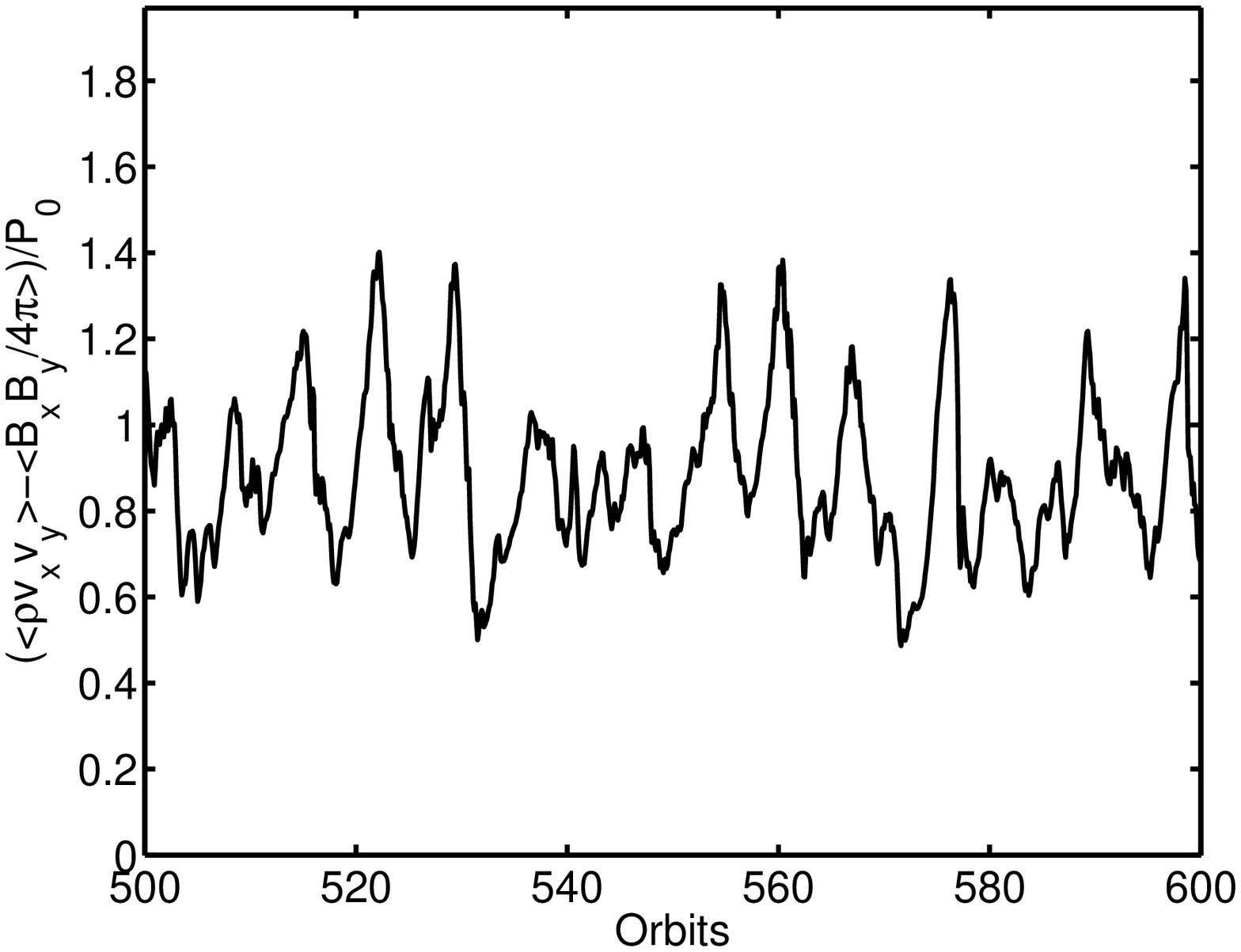}
  \leavevmode\epsfxsize=6.0cm\epsfbox{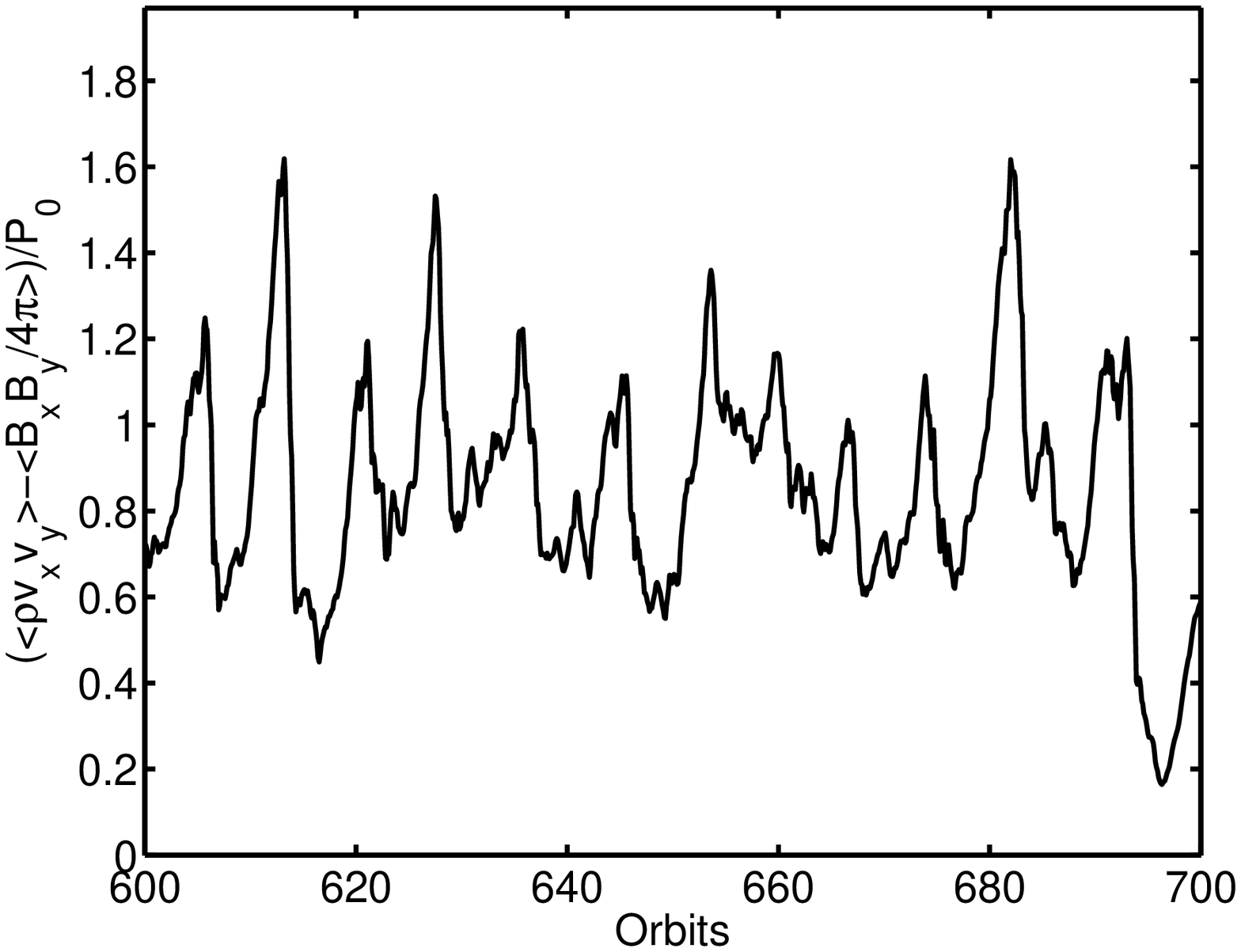}
  \leavevmode\epsfxsize=6.0cm\epsfbox{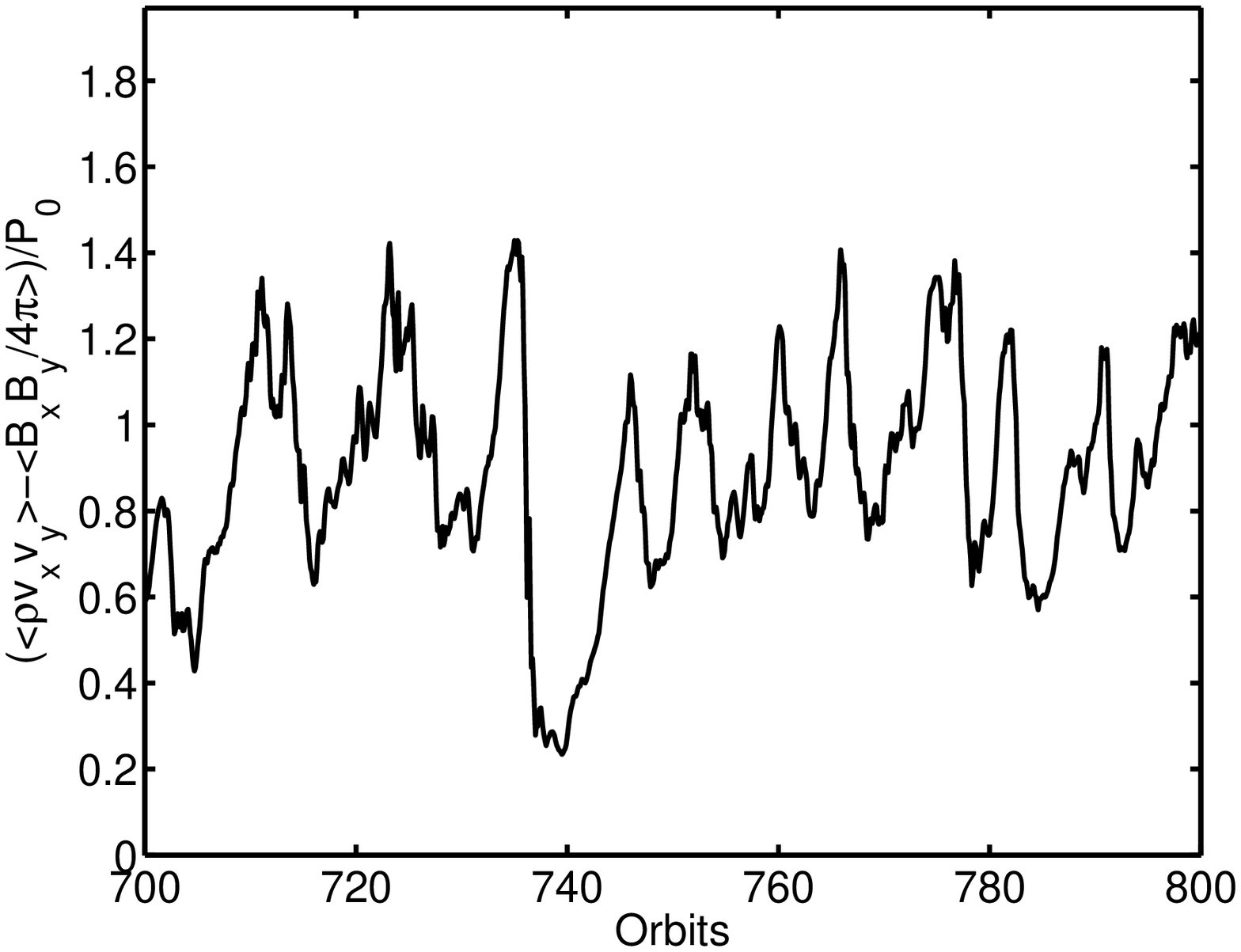}
  \leavevmode\epsfxsize=6.0cm\epsfbox{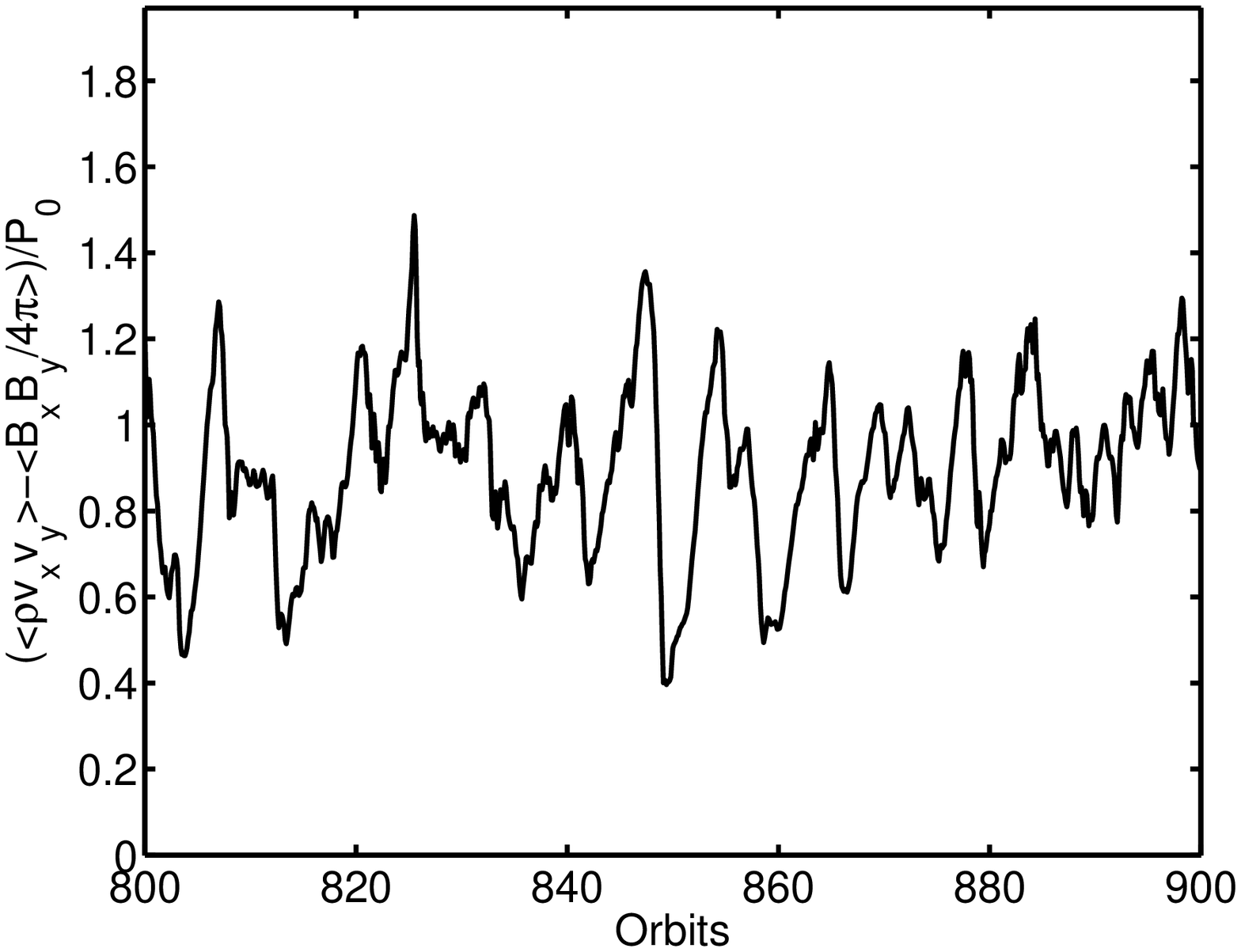}
  \leavevmode\epsfxsize=6.0cm\epsfbox{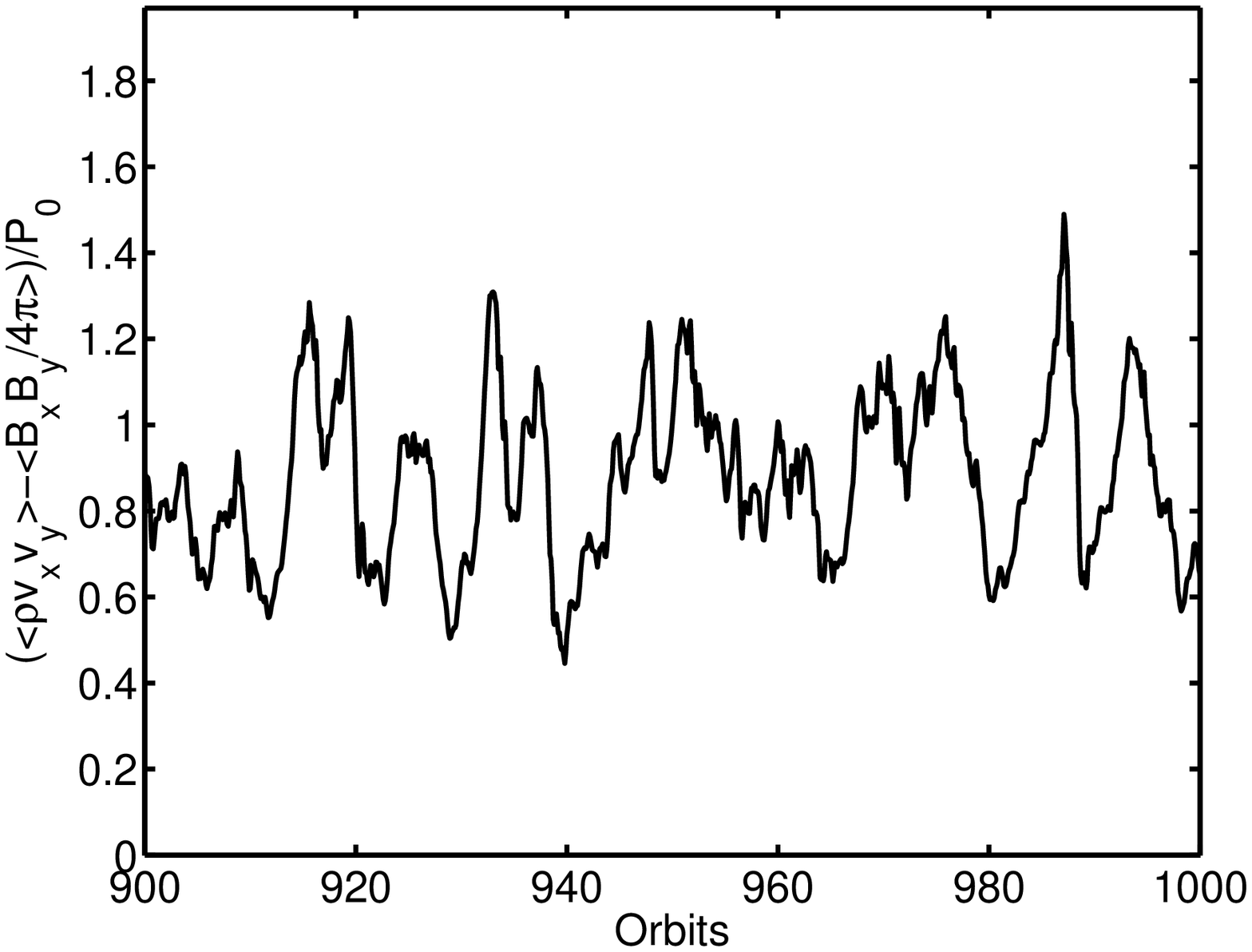}
    \caption{A series of figures showing the time evolution of
    $\alpha$ when 'cleaning' is employed for case 6.}\label{Removesnaps}
\end{center}
\end{figure*}

Figure \ref{remove1} shows the new temporal evolution of $\alpha$
once the cleaning has been carried out for the $32^3$, which is case 6. It is clear that the effect of cleaning the mean values on
$\alpha$ is to continue with the saturated level that we witnessed
in the first part of figure \ref{fiden2}. Therefore, it is now
possible to evaluate the cumulative average until it saturates. Figure
\ref{remove1} not only shows the cumulative average but also the mean
value for this case. Figure \ref{Removesnaps} shows a more detailed
series of snapshots for $\alpha$ in this case and clearly shows that
the randomised, turbulent, behaviour persists for all time and we do
not now move to a repeated channel solution case.

\begin{figure}
\begin{center}
    \leavevmode\epsfxsize=8cm\epsfbox{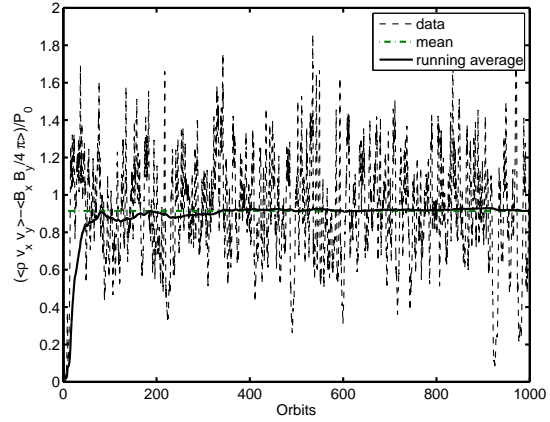}
    \caption{The evolution of $\alpha$ when full cleaning is employed
    for case 7 (1:1:1 aspect ratio)}
\label{remove2}
\end{center}
\end{figure}

The primary aim of this paper was to examine the effect of
decreasing the dissipation scale (by increasing the resolution) and
we are now in a position to do so. Figure \ref{remove2} show the
effect of doubling the  resolution (decreasing the dissipation
scale) on $\alpha$. We note that the average value is increased as
resolution is increased. This fact is amplified by the data in table
\ref{table2} (cases 6-8), which shows a clear upward trend as
resolution is increased. As such, this shows that small scales are
important in these calculations and the level at which the
dissipation scale is set is crucial is determining that rate of
enhanced transport.

\begin{table}
\begin{center}
 \centering
  \caption{Selected cases together and the associated $\alpha$ from a cumulative average}\label{table2}
  \begin{tabular}{@{}cccc@{}}
  \hline
 Case & Resolution & Box dimension & $\alpha$  \\
 \hline

6& 32:32:32& 1:1:1& 0.85\\
7& 64:64:64& 1:1:1& 0.94\\
8& 128:128:128& 1:1:1&  1.02\\
9& 64:128:64& 1:2:1& 0.43\\
10& 64:192:64& 1:3:1&  0.29\\
11& 64:384:64& 1:6:1&  0.16\\

\hline
\end{tabular}
\end{center}
\end{table}

\subsection{The Effect of Increasing the Box length}

\begin{figure}
\begin{center}
    \leavevmode\epsfxsize=8cm\epsfbox{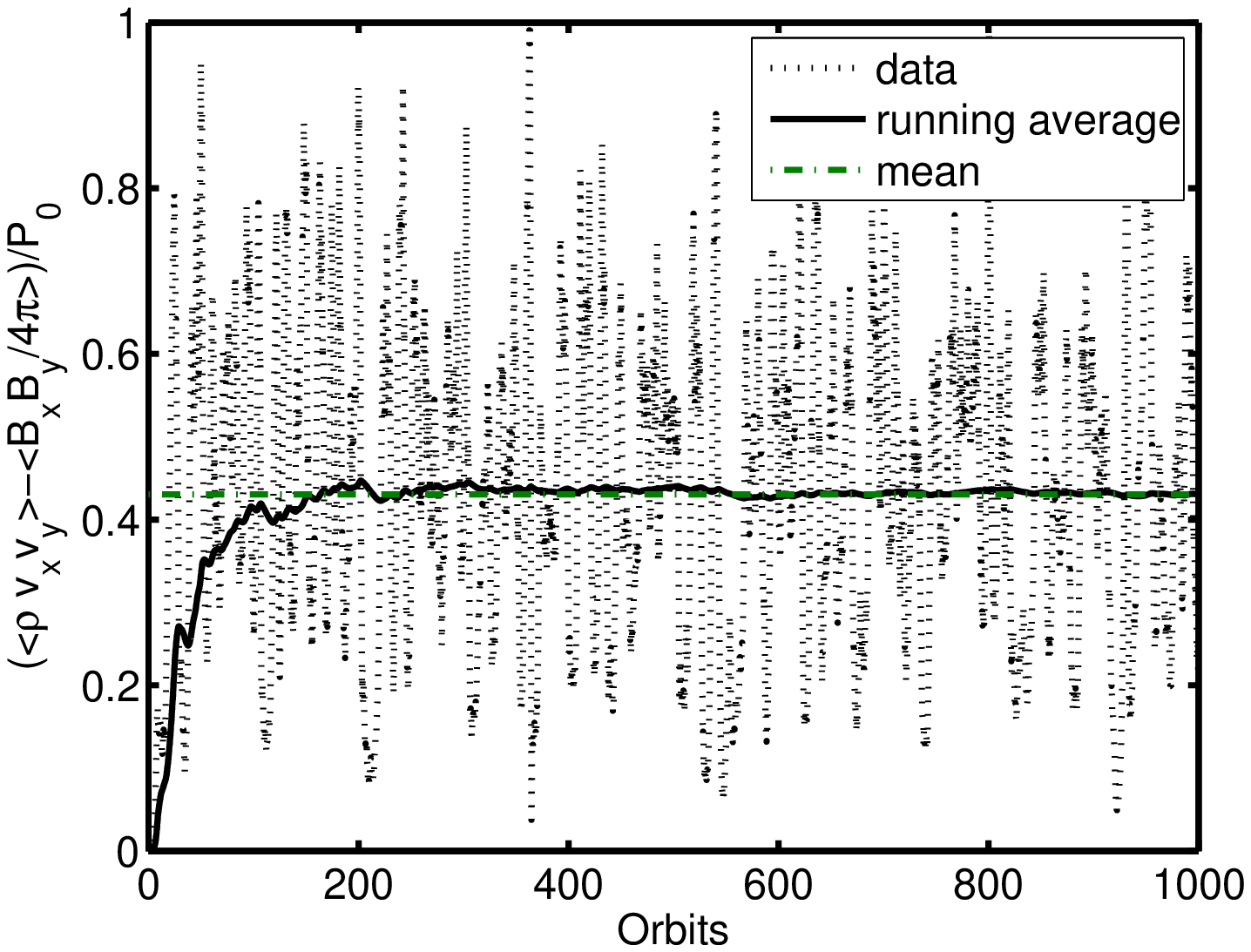}
    \leavevmode\epsfxsize=7.2cm\epsfbox{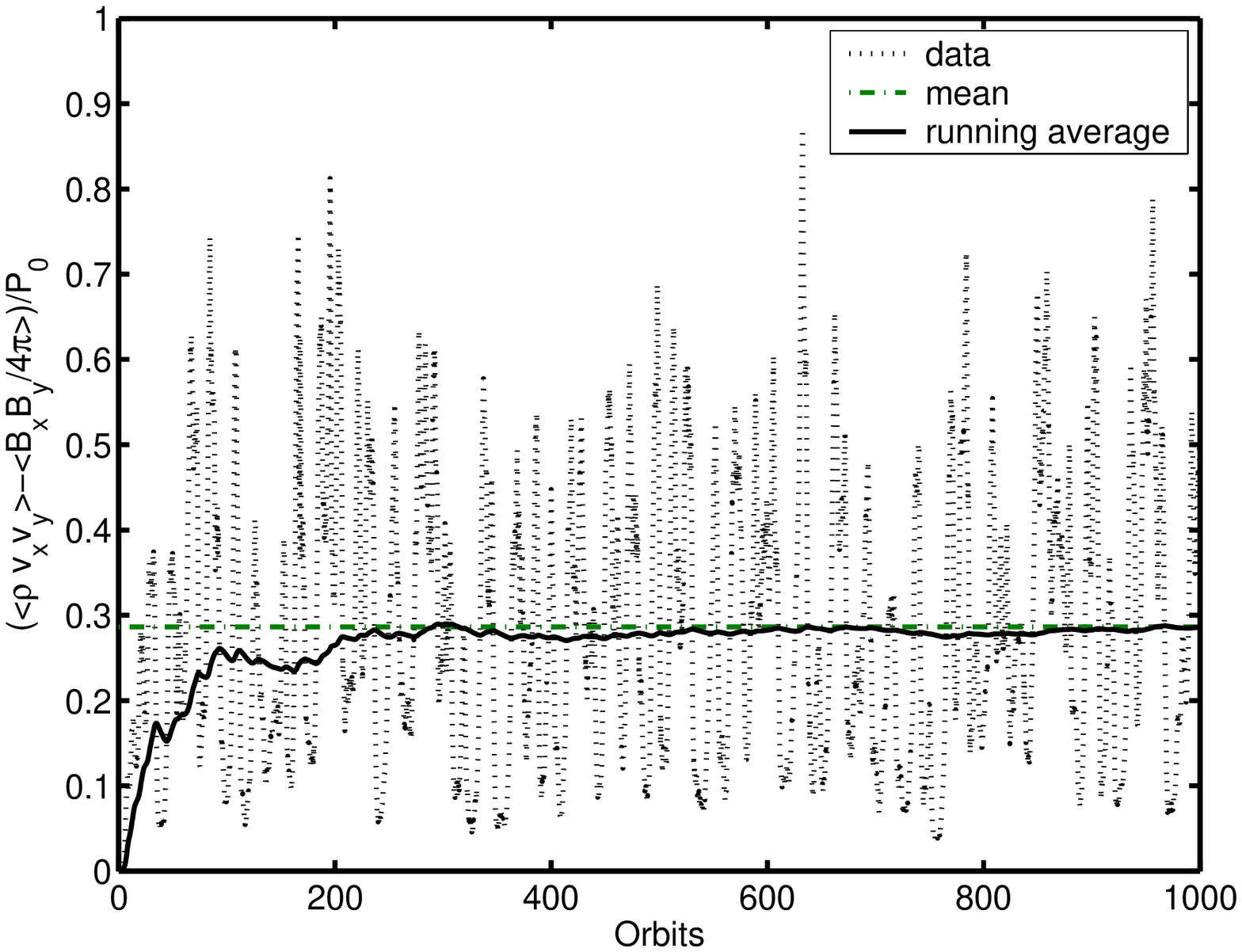}
   \leavevmode\epsfxsize=8cm\epsfbox{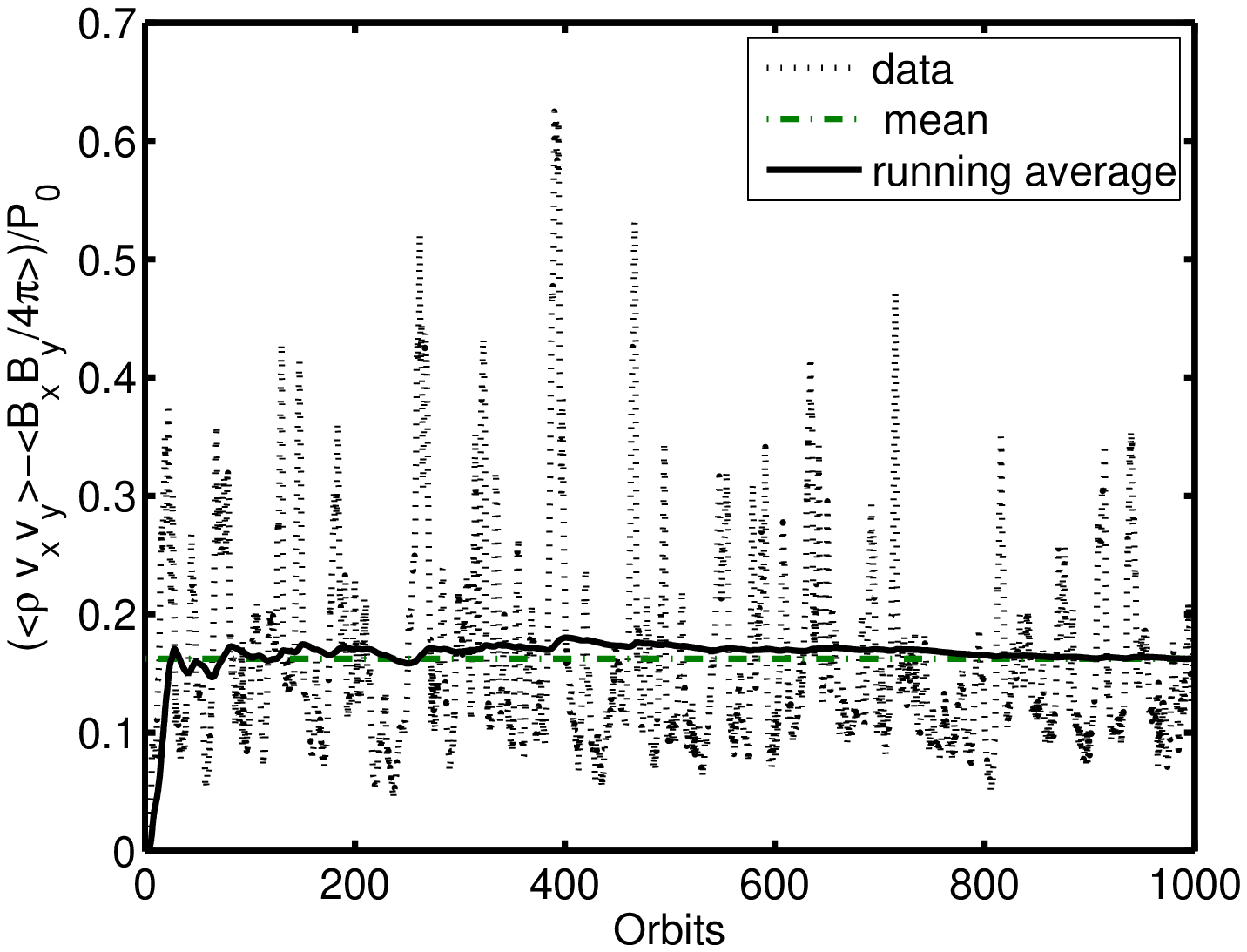}
    \caption{Top: The evolution of $\alpha$ when full cleaning is
    employed for case 9 (1:2:1 aspect ratio). Middle:The evolution of $\alpha$ when full cleaning is
    employed for case 10 (1:3:1 aspect ratio). Bottom: The evolution of $\alpha$ when full cleaning is
    employed for case 11 (1:6:1 aspect ratio) } \label{121}
\end{center}
\end{figure}

In the above section we chose to focus on a domain where the aspect
ratio is $1:1:1$. This was selected to facilitate the longer time
calculations at higher resolutions. However, as the earlier
calculations considered a $1:2\pi:1$ and so in this section we
examine what happens at modest resolutions to the saturated level of
$\alpha$. We have considered boxes with the aspect ratios $1:2:1$,
$1:3:1$ and $1:6:1$. The effect of elongating the is summarised in
\ref{table2} (cases 7, 9-11) and the evolution is shown in figure
\ref{121}.

There is a clear decrease in the saturation level as the box is
increased in the azimuthal direction. What we are seeing is the
effect that in short box lengths the parasitic instability is
limited (see \cite{GX} for more details on this instability).

\section{Conclusions}

In this paper we have returned to the original concept to understand
turbulent transport of angular momentum in an accretion disc. As
such, we considered the standard local shearing box model in which
the ideal MHD equations were solved. We chose to start by
considering a cube with sides of equal length because we wished to
have large integration times (1000 orbits) and a numerical
dissipation that the same in all directions.

Our calculations shown that, without correction, there is a shift in
the accretion rate (a shift in the turbulent transport rate). Such a
shift is interesting due to the fact that there are shifts in
spectra of accretion discs, which imply a corresponding shift in
accretion rate. However, in this case, the shift is unintentional
and a by-product of the interpolation in the shearing box boundary
conditions that introduce errors because of the interpolation that
is a necessary part of the procedure. This error is unavoidable in
such calculations and does remain small for low numbers of orbits.
However, for large numbers of orbits the error in shown to compound
to such an extent that the mean $B_y$ field, which should be zero,
grows and becomes larger in magnitude that the imposed $B_z$ field.
This error filers into the various components of both the velocity
and magnetic fields.

In this paper we sought to consider the simple case of an
imposed vertical magnetic field i.e.\ without  a shift to a state
where where was a mean field in more than one direction. As such we
applied a simple fix to this issue that removes the error that is
induced into the mean value of each of the components of the
velocity and magnetic fields. This facilitated us to evaluate a
long-term cumulative average for $\alpha$, which is a better quantity
when you wish to compare $\alpha$ values for different parameters.
Following this procedure, we showed that increasing the resolution,
for a fixed box size, gives rise to an increase in the value of
$\alpha$, which suggests that the values here are probably a lower
bound for $\alpha$ values in these kind of simulations. This leads us to deduce that small scales are playing and
important role in determining the rate of accretion. In an accretion
disc the dissipative length-scale is much less than we considered
here and so we conclude that in a disc there will be a significant
enhancement to the rate of angular momentum transport.

As mentioned above, the work in the first part of this paper is
not in the standard domain of the early work in a shearing box i.e.\
it was conducted in a cube with sides of equal length. This was
selected to facilitate our desire for higher resolutions and and
extremely long integration times. However, it of obvious interest to
examine the effect of increasing the box length at modest
resolutions. We found that increasing the box length led to a
reduction in the mean value of $\alpha$. This is attributed to the
increased freedom for the parasitic instability as the box length is
increased (see \cite{GX} for a detailed discussion of this
instability). We note here that while that increasing the resolution
by a factor of four leads to a modest (15
box length is much more profound and increasing the box length by a
factor of three leads to a reduction of $\alpha$ such that it is almost
$2/3rds$ of its value for a box with length 1 in the y-direction.

While we acknowledge that there is still considerable working to be
conducted in this area to fully understand the magnetorotational
instability we believe that this paper is an important contribution
to the field as it clearly demonstrates the need for great care in
numerical calculations. Further it shows that, for the case of a
imposed vertical magnetic field, $\alpha$ is non-negligible and it
points to there existing sufficient transportation within a disc to
give accretion in an appropriate time-scale.

The reader should note that what has been found here is for the case
of a uniform vertical magnetic field and that this is very
different to the case of a zero-mean magnetic field as considered
by, for example, \cite{P2007} \cite{FP1} and \cite{FP2}, where $\alpha$ in many
situations was found to be extremely small. Here we have shown that
$\alpha$ increases as the scale of dissipation decreases and that it
decreases as the length of the computational domain in the azimuthal
direction is increased. It is not clear to what extend zero-mean flux
calculations for long times are effected by the error that we found
and we would suggest that all calculations in a shearing box check
that mean quantities are maintained to expected values.

This paper does pose some questions of which the most obvious is
whether or not a change in the balance of mean components of the
magnetic field, along with the strength of the magnetic field,
always gives rise to a change in the accretion rates. This issue is
being investigated at present.

\section*{Acknowledgments}

We would like to thank Steve Balbus, John Hawley, Pierre Lesaffre
and Jim Stone for suggestions and comments on during this work. We
wish to thank the Ecole Normale Superieure for the supercomputer
known as JxB on which these calculations were carried out. We also
wish to thank the referee for helpful suggestions.

\label{lastpage}

\end{document}